\documentclass[final]{elsarticle}
\pdfoutput=1
\usepackage{amsmath}
\usepackage{amssymb}
\usepackage{graphics,graphicx}
\usepackage{epstopdf}
\usepackage{hyperref}
\usepackage{color}
\usepackage{caption}
\usepackage{subcaption}
\usepackage{todonotes} 

\makeatletter
\def\ps@pprintTitle{%
 \let\@oddhead\@empty
 \let\@evenhead\@empty
 \def\@oddfoot{}%
 \let\@evenfoot\@oddfoot}
\makeatother

\newcommand{\fref}[1]{Fig.~\ref{#1}}

\begin{document}

\begin{frontmatter}

\title{Stress in a polymer brush}

\author[mech]{M. Manav}
\ead{manav@alumni.ubc.ca}

\author[mech]{M. Ponga}
\ead{mponga@mech.ubc.ca}

\author[mech]{A. Srikantha Phani\corref{cor1}}
\ead{srikanth@mech.ubc.ca}

\cortext[cor1]{Corresponding author}

\address[mech]{Mechanical engineering, University of British Columbia, Vancouver, BC V6T 1Z4, Canada}

\begin{abstract}
We study the stress distribution in a polymer brush material over a range of graft densities using molecular dynamics (MD) simulations and theory.  Flexible polymer chains are treated as  beads connected by nonlinear springs governed by a modified finitely extensible nonlinear elastic (FENE) potential in MD simulations.  Simulations confirmed the quartic variation of the normal stress parallel to substrate, within the bulk of the brush, as predicted in our previous  work, for low graft densities. However, in the high graft density regime, the Gaussian chain elasticity assumption is violated by finite extensibility effects (force-extension divergence) and the restriction to binary interaction among monomers is insufficient.  This motivated us to extend a semi-analytical strong stretching mean field theory (SST) for polymer brushes, based on Langevin chains and a modified Carnahan-Starling equation of state to model monomer interactions. Our extended theory elucidates the stress and  monomer density profiles obtained from  MD simulations, as well as reproduces Gaussian chain results for small graft densities. A good agreement is observed between predictions of  MD and Langevin chain SST for monomer density profile, end density profile and stress profile in high graft density regime, without fitting parameters (virial coefficients). Quantitative comparisons of MD results with various available theories suggest that excluded volume correlations may be important.
\end{abstract}


\end{frontmatter}

\section{Introduction}
Long polymer chains end-grafted  on an  impermeable substrate stretch away, in the presence of a good solvent\footnote{Here a good solvent condition means high affinity of monomers for solvent molecules while in a poor solvent condition, monomers minimize exposure to solvent molecules. In a $\theta$-solvent condition, there is no binary interaction between monomers.}, forming a polymer brush (PB) structure~\citep{alexander77,de80,milner91,edmondson2004}.  Polymer brushes have been shown to deform the substrate reversibly and controllably, in response to an external stimuli such as a change in temperature, pH, light etc.~\citep{zou2011}, opening a new class of soft active matter (SAM). Surface modification induced stimulus response, a facile technology, has a distinct advantage over other SAMs that require bulk modification, as in stimuli-responsive hydrogels, electroactive polymers, liquid crystal elastomers etc.  PB-SAMs have found use in many technologies: as a programmable material~\citep{kelby2011},  in sensing and actuation~\citep{abu2006micro,klushin2014}, as microcantilever coating in glucose sensing~\citep{chen2010} and selective ion sensing~\citep{peng2017}, microcantilever actuation~\citep{zhou2006,zhou2008}, and macroscale bending stretching actuation involving large substrate deformations~\citep{zou2011}. An overview of a polymer brush and its applications can be found in~\citep{stuart2010} and~\citep{azzaroni2012}.

This work builds on an earlier study~\cite{manav2018} in two respects: numerical simulations and extension of theory. We report MD simulations aimed to understand stress variation within a brush as  a function of its molecular parameters. Second, we extend stress expressions in~\cite{manav2018} into high graft density regimes. We begin this paper with an overview of various theories in Section~\ref{overview}, from the perspective of mechanical stress in polymer brushes.  We then extend strong stretching theory (SST) for brushes~\citep{milner88,skvortsov88,zhulina91} with non-Gaussian chains~\cite{shim89,amoskov94,biesheuvel2008} to calculate stress in the brush. Using Langevin chain elasticity and a modified Carnahan-Starling (CS) equation of state~\cite{biesheuvel2008}, we find stress distribution in a densely grafted polymer brush in a good solvent in Section~\ref{SurfS_pb}. This particular choice  enables the derivation of (semi)analytical expression for stress,  free of fitting-parameters, to cover a wide range of graft densities studied in our MD simulations in Section~\ref{MD}, albeit under good solvent conditions\footnote{A Flory-Huggins equation of state or an enthalpic correction to CS equation of state can be considered to include different solvent qualities.}. Predictions by various theories are compared with MD simulation results and discussed in Section~\ref{result}, ending with concluding remarks in Section~\ref{conclusion}.

\section{An overview of polymer brush theories}\label{overview}
 The structure of a PB results from excluded volume interactions generated by solvent molecules surrounding monomers, entropic resistance to stretching of polymer chains, and the constraint imposed by end grafting. While excluded volume repulsion among monomers makes a polymer chain stretch away from the grafting surface, the entropic spring effect brings the two ends of a chain together,  and their resulting balance dictates the brush formation.  Mechanistically, a brush grafted to a substrate can be seen as an elastic surface layer with stress \cite{begley2005,utz08,manav2018}, which deforms the elastic substrate it is grafted to~\citep{zou2011,manav2018}. A stimulus modifies the excluded volume interaction, leading to a change in the brush structure as well as the stresses and the surface elastic properties of the brush, thus offering unprecedented control on the substrate deformations~\citep{zou2011,manav2018}.

For several decades, polymer brushes have been a system of immense interest to polymer physicists. Multiple, often complementary, theoretical approaches have emerged to relate the macroscopic brush properties such as brush height ($H$) to molecular scale parameters such as effective monomer size ($a$), number of monomers in a chain ($N$), graft density\footnote{Number of chains grafted to unit area of a substrate.} ($\rho_g$) etc. of a brush.  A detailed comparative review can be found in~\citep{halperin94,netz03,binder12}. Here, we recall salient features to serve as a useful background to this study and motivate the reader to explore these theories in detail. We note that the above theories do not focus on the variation of mechanical stress within a brush but are limited to macroscopic brush properties.

Alexander~\cite{alexander77}  propounded a scaling theory by invoking Flory-like mean field argument under the assumptions of: (a)  ideal chain (Gaussian chain) statistics and (b) that \emph{all} end-points of polymer chains are at a constant height $H$ above the grafting surface, the so called \emph{step} profile \emph{ansatz} for monomer density. Minimization of free energy, a sum of stretching free energy and interaction free energy, gives the scaling relation $H\sim\rho_g^{1/3}$. Drawing on the theory of critical phenomena, de Gennes constructed a scaling theory~\cite{de80}, using  blobs of different sizes associated with different energy scales in a brush system. An excellent review of \emph{blobology} based scaling arguments can be found in~\citep{halperin94}. Although, the blob theory accounts for self-avoiding random walks of individual polymer chains, it still has the step profile ansatz of Alexander~\cite{alexander77}. This is relaxed in a mean field theory~\cite{dolan74,scheutjens79} for a brush, by placing  a polymer chain in a position-dependent effective mean field potential which is dependent on local monomer density, thus accounting for the influence of neighboring chains. Fluctuations in the interaction field of a chain with the surrounding chains is ignored in fully-numerical lattice based calculations~\citep{dolan75,scheutjens79}, with no assumptions made on monomer density profiles. The fact that the effective mean field potential and the monomer density at the minimum free energy configuration of the brush are self consistent, hence the name self consistent field theory (SCFT)~\citep{cosgrove87,milner90}, is used to obtain density profile numerically. Unlike scaling theory, however, mean field theory predicts that the free ends of polymer chains in a brush are distributed throughout the brush and also that the monomer density profile need not be a step function. 

A departure from these lattice-based numerical calculation is the recognition that, starting from any free-end, polymer chains follow  \emph{classical} paths, provided the  brush is strongly stretched. In this strong stretching regime, the classical paths of polymer chains dominate the partition function of the brush and fluctuations from these paths can be ignored~\citep{semenov85}. This crucial insight allowed the development of the so called strong stretching theory (SST) for brushes~\citep{milner88,skvortsov88,zhulina91}. It must, however, be noted that the entropic elasticity of the chain was taken to be entirely Gaussian in these theories initially, without accounting for any divergence in the force-extension relation~\cite{rubinstein2003}. Further, in a \emph{moderately} dense brush binary interactions among monomers are dominant, which tantamount to the truncation of virial expansion. With these two restrictions in the calculation of interaction free energy, and stretching free energy, a \emph{parabolic} monomer density profile was predicted analytically~\citep{milner88,skvortsov88}. We denote this theory as SST-Gaussian, or SST-G for short. The parabolic monomer density profile is at odds with earlier scaling theories, and later confirmed to be correct by rigorous MD simulation studies~\citep{murat89,grest93,dimitrov07} and by experiments~\citep{auroy92,karim94}.  Deviations were recognized at the grafted and free ends of a brush due to a depletion layer and a tail, respectively. Furthermore, the chain ends are assumed to be stretch free in SST, but they have been observed to undergo different end-stretching depending on their location from the grafting surface~\cite{Siedel}. When the free ends of the chain are far away from the grafting surface, they point \emph{away} from the surface, while those close to the surface point \emph{toward} the surface, as observed in Monte Carlo (MC) and MD simulations, however mean end stretching is found to be zero as assumed in SST~\cite{binder95,Siedel}. A further refinement of SST-G theory is made by accounting for the force-extension divergence~\cite{shim89} using Langevin chain elasticity in~\cite{amoskov94}. A (semi) analytical procedure emerges which can predict monomer density profiles over a range of graft densities, which smoothly bridge the parabolic and near-step profiles. Further, a fitting-parameter free procedure follows by replacing the original Flory-Huggins (FH) equation of state in~\cite{amoskov94} with a modified Carnahan-Starling equation of state with a correction for connection between beads in a polymer chain, for a good solvent~\cite{biesheuvel2008}. The restriction on solvent quality can be relaxed by reverting to FH equation of state or correcting the CS by adding an enthalpic term~\cite{biesheuvel2008,romeis2012}.

End to end distance of a chain in a good solvent is large at high grafting densities, and hence the force-extension divergence~\cite{rubinstein2003}, absent in Gaussian elasticity, must be considered in elastic free energy calculation. A dimensionless extension parameter, $\beta_e$, defined as the ratio of the end to end distance ($\sim H$) and contour length of a chain ($Na$): $\beta_e=H/(Na)$, can be used as a gauge. When $\beta_e>1/3$, divergence in force-extension due to finite extensibility  cannot be ignored. We note that neither Alexander-de Gennes scaling nor SST-G accounts for force-extension divergence. A semi-analytical framework to account for finite extensibility effects in SST was proposed in~\citep{shim89}. However, the form of stretching free energy was chosen based on mathematical convenience. Semi-analytical SST using stretching free energy of a Langevin chain, which accurately describes large stretching of a freely jointed chain, was developed in~\citep{amoskov94}. It predicts that, with increasing density, a brush approaches a step profile for monomer density as suggested by scaling theory and the chain free ends increasingly straddle the free surface of the brush. However, the mean field potential in this work was obtained as a series solution. A rational polynomial approximation for the series solution is proposed in~\citep{biesheuvel2008}. Furthermore, and conventionally, interaction free energy in SST is calculated based on FH theory. CS equation-of-state from liquid-state theory for hard-sphere mixtures with correction for connection between beads in a polymer chain to account for interaction is employed in~\citep{biesheuvel2008,romeis2012}. Remarkably, prediction of brush structure from this SST shows close match with the bead-spring MD simulation results without any need for a fitting parameter~\citep{biesheuvel2008,romeis2012}. Here, we extend this theory to calculate stress in a brush.  Henceforth, this refinement of SST-G based on Langevin chain elasticity is referred to as SST-L, without delineating the equation of state.

While numerical mean field calculations show good match with SST-G for graft densities $\rho_g<0.2$~\cite{milner90}, these calculations ignore fluctuations from the effective mean field. Molecular scale simulation methods~\cite{binder95} such as MC simulation~\cite{chakrabarti90,lai91,laradji94} and MD simulations~\cite{murat89,grest93} account for such fluctuations. These simulations show that the brush height scaling predicted by the analytical theories hold for a limited range of graft densities and large $N$. So, a molecular scale simulation not only validates the analytical theories but will also reveal the validity of their assumptions in terms of the range of molecular scale parameters. A bead-spring model of a polymer chain was pioneered in~\cite{grest86}. In this model, each polymer chain is represented by a series of connected beads. The beads represent effective monomers. The interaction between connected and unconnected beads are governed by different potentials in an MD simulation. The model was used to simulate polymer brushes in different solvent conditions~\cite{murat89,grest93}. MD studies of moderate and high graft density brushes have been reported in~\cite{he2007}. A comparison between static properties of a brush in a good solvent obtained from different models can be found in~\cite{kreer04}. In this work, we use a bead-spring model for polymer chains to obtain stresses. We investigate the behavior of PBs of different graft densities, and use MD simulations to assess theories. A limitation of MD simulations is that in order to achieve strong stretching of chains to be able to compare MD with SSTs, one requires a large number of beads per chain, and this number increases considerably for low graft densities. Unfortunately, this makes it exceedingly expensive to equilibrate the system, and one seeks a reasonable trade off between accuracy and efficiency. We will see later (Section \ref{result}) that this has implications in the prediction of mechanical stress. For completeness, Table~\ref{limitation_SST} compares the main assumptions and features of SST-G, SST-L and MD.  

\begin{table}[!h]
	\caption{\label{limitation_SST} A comparison of the assumptions and limitations of SST and MD.}
	\center{
	\begin{tabular}{llll}
		\hline
		  Feature& SST-G & SST-L & MD \\ 
		   \hline
		 Chain elasticity & Gaussian chain & Langevin chain & No limitation \\  
		  Stretching&Infinite&Infinite&Finite\\
		 Self-avoidance & Ideal chain & Ideal chain & Considered \\  
		 Virial truncation &  Binary & No truncation & No truncation\\
		 Chain paths & Classical & Classical & All paths \\  
		 Chain end & Force free & Force free & Unconstrained \\  
		 Computational cost & None & Small & Very high \\ \hline
		
	\end{tabular}}
\end{table}

%%%%%%%%%%%%%%%%%%%%%%%%%%%%%%%%%%%%%%%%%%%%%%
\section{Stress in a polymer brush using SST}
\label{SurfS_pb}
\begin{figure}[!h]
\center
\includegraphics[width=10cm]{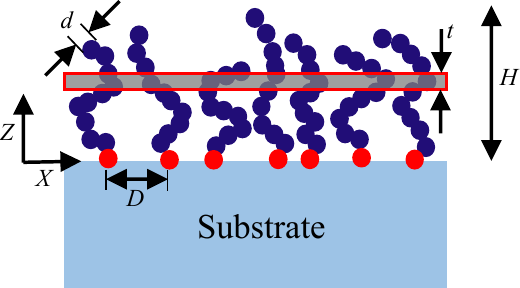}
\caption{A planar polymer brush of height $H$, effective monomer diameter $d$, effective length of a monomer $a$ and contour length of a polymer chain $Na$. Typically, $d=a$. Inverse square root of graft density equals the average distance between grafting points in the brush ($\langle D\rangle=\rho_g^{-1/2}$). A thin layer at height $\textsc{z}$ is also shown that we frequently refer to in the text.}
\label{brush_slit}
\end{figure} 
Consider a \emph{neutral} brush of graft density $\rho_g$,  with $N$ monomers in each chain (monodisperse brush) grafted on a \emph{rigid} substrate, as sketched in the schematic in~\fref{brush_slit}, extending to a height $H$. Introduce a co-ordinate, $\textsc{z}$, measured with respect to the grafting surface as the datum $\textsc{z}=0$. Then, the mean field ($V(\textsc{z})$) is dependent on local monomer density ($\phi(\textsc{z})$), which in turn governs the chemical potential $\mu(\phi)$, thus giving~\cite{shim89,amoskov94,biesheuvel2008}:
\begin{equation}
V(H)-V(\textsc{z})=\mu(\phi(\textsc{z}))-\mu(\phi(H)). \label{eq2p0}
\end{equation}
In a brush in good solvent, monomer density is $0$ at $\textsc{z}=H$. Hence, usually $\mu(\phi(H))$ is also assumed to be $0$.  The chain ends are distributed throughout the brush and the distribution function $g(\textsc{z})$ obeys $\int_0^Hg(\textsc{z})=\rho_g$.

The brush structure results from the competition between monomer-monomer interaction and stretching of the brush. So, free energy per unit substrate area of a brush, $F$, is the sum of interaction free energy $F_{int}$ and chain stretching free energy $F_{el}$. 
\begin{equation}
F=F_{int}+F_{el}. \label{eq2p1}
\end{equation}
In a strong stretching mean field description of a brush, where each chain is assumed to follow a minimum energy path away from the grafting surface, free energy density $f(\textsc{z})$ (free energy per unit volume)  within the brush can also be obtained. Then, free energy of the brush can be written as:
\begin{equation}
F=\int_0^H f(\textsc{z}) d\textsc{z}=\int_0^H f_{int}(\textsc{z}) d\textsc{z} +\int_0^H f_{el}(\textsc{z}) d\textsc{z}, \label{eq2p2}
\end{equation}
where $f_{int}(\textsc{z})$ and $f_{el}(\textsc{z})$ are interaction and elastic free energy densities, respectively. The free energy density in a brush is nonuniform. By applying an infinitesimally small uniform horizontal strain $\epsilon_{\textsc{xx}}$ to the brush and calculating the change in free energy density in the brush, with the assumption of plane strain in $\textsc{y}$-direction, and that all the shear stresses as well as normal stress in the $\textsc{z}$-direction are zero, stress distribution within the brush can be obtained. It is shown in \cite{manav2018} that stress then is given by:
\begin{equation}
\sigma_{\textsc{xx}}=\frac{\partial f(\textsc{z})}{\partial\epsilon_{\textsc{xx}}}+f(\textsc{z})\left(1+\frac{\partial\epsilon_{\textsc{zz}}}{\partial\epsilon_{\textsc{xx}}}\right),
\label{eq2p3}
\end{equation}
where $\epsilon_{\textsc{zz}}$ is the infinitesimal normal strain in the $\textsc{z}$-direction due to strain $\epsilon_{\textsc{xx}}$ applied in the $\textsc{x}$-direction. When the extension parameter $\beta_e\le1/3$, the divergence-free Gaussian elasticity is reasonable, and stress calculations in this regime, fully derived in an earlier work~\cite{manav2018}, are recalled first. Then, we will  present a (semi) analytical extension of stress expressions based on SST with Langevin chains (SST-L) in Section~\ref{SST_Langevin}. 

\subsection{SST with Gaussian chains (SST-G)}
\label{SST_Gaussian}
We calculate $ f_{el}(\textsc{z})$ and $ f_{int}(\textsc{z})$ based on Gaussian elasticity and FH solution theory~\cite{milner88,zhulina91,manav2018}, respectively. In a \emph{moderately} dense brush, binary interaction dominates. Hence, the chemical potential, $\mu(\phi)$, is related to the second virial coefficient, $v$, and monomer density, $\phi(\textsc{z})$, via:
\begin{equation}
\mu(\phi)=v\phi(\textsc{z}). \label{eq2p3a}
\end{equation}
With the above truncation due to the restrictions placed on the monomer interactions,  the mean field potential is then obtained in~\cite{milner88,zhulina91} as:
\begin{equation}
V(\textsc{z})=\frac{3\pi^2}{8N^2a^2}\textsc{z}^2. \label{eq2p3b}
\end{equation}
Using~\eqref{eq2p3a} and~\eqref{eq2p3b} in~\eqref{eq2p0}, monomer density, $\phi$ can be evaluated. Defining $E(\textsc{z},\zeta)$ as local stretching at height $\textsc{z}$ in a chain with end at height $\zeta$, free energy density in the brush is~\cite{manav2018}:
\begin{equation}
f(\textsc{z})=\underbrace{\frac{1}{2}v\phi^2(\textsc{z})k_BT}_{f_{int}}+\underbrace{\frac{3}{2a^2}k_BT\int_{\textsc{z}}^{H}g(\zeta)E(\textsc{z},\zeta)d\zeta}_{f_{el}}, \label{eq2p4}
\end{equation}
where $k_B$ and $T$ are Boltzmann constant and absolute temperature, respectively. SST-G provides the following expressions for $\phi(\textsc{z})$, $H$, $g(\zeta)$, and $E(\textsc{z},\zeta)$ \cite{milner88,zhulina91}.
\begin{eqnarray}
\phi(\textsc{z})=\frac{3\pi^2}{8N^2va^2}\left(H^2-\textsc{z}^2\right),\label{eq2p5a}\\
H=\left(\frac{4}{\pi^2}\right)^{1/3}v^{1/3}a^{2/3}\rho_g^{1/3}N, \label{eq2p5b} \\
g(\zeta)=\gamma\zeta\sqrt{H^2-\zeta^2}, \quad \gamma=\frac{3\pi^2}{4N^3va^2}, \label{eq2p5c} \\
E(\textsc{z},\zeta)=\frac{\pi}{2N}\sqrt{\zeta^2-\textsc{z}^2}. \label{eq2p5d}
\end{eqnarray}
Noting that $\frac{\partial\epsilon_{\textsc{zz}}}{\partial\epsilon_{\textsc{xx}}}=-\frac{\textsc{1}}{3}$, and evaluating $\frac{\partial f(\textsc{z})}{\partial\epsilon_{\textsc{xx}}}$, the stress distribution in a brush was obtained by using \eqref{eq2p3} in \cite{manav2018}.
\begin{equation}
\sigma_{\textsc{xx}}=-\frac{9}{8}\left(\frac{\pi^2}{12}\right)^{2/3}v^{1/3}\rho_g^{4/3}\beta^{2/3}\left(1-\left(\frac{\textsc{z}}{H}\right)^2\right)^2k_BT.
\label{eq2p6}
\end{equation}
Two important conclusions emerge in the Gaussian elasticity setting. First,  a quartic variation of stress with respect to distance from the grafting surface with the maximum at the grafting surface. Second, a strong dependence on graft density ($\rho_g$) compared to number of effective monomers ($N$) in a polymer chain. In this work, we seek to validate these theoretical predictions for stress with MD simulation results, which are free from the assumptions of SST-G (see Table~\ref{limitation_SST}). 

We now consider the case of finite chain extensibility, where $\beta_e>1/3$, and force extension divergence, a ``hardening" entropic spring characteristic, is important.

\subsection{SST with Langevin chains (STT-L)}
\label{SST_Langevin}
\begin{figure}[!h]
\center
\includegraphics[width=8.7cm]{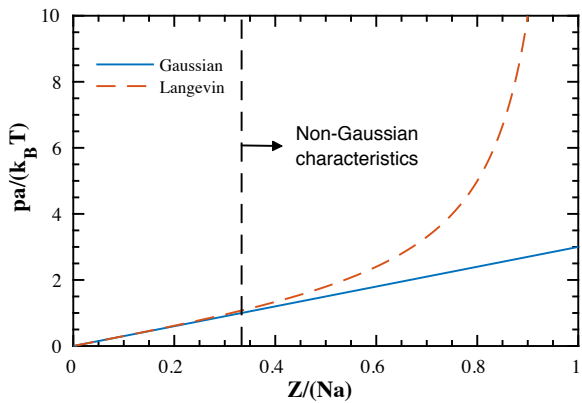}
\caption{Force-extension curves for an isolated Langevin chain and the corresponding isolated Gaussian chain. Note the force-extension divergence at higher stretching giving a ``hardening" spring. For small extensions, the two curves are indistinguishable.}
\label{force_extension}
\end{figure}

The average distance $\textsc{z}$ between ends of a Langevin chain with $N$ effective monomers of length $a$, due to a force $p$ at the ends is given by Langevin function ($\mathcal{L}(\cdot)$)~\cite{rubinstein2003}:
\begin{equation}
\frac{\textsc{z}}{Na}=\mathcal{L}\left( \frac{p a}{k_BT}\right), \quad \mathcal{L}(\cdot):=\coth (\cdot)-\frac{1}{(\cdot)}. \label{eq2p7}
\end{equation}
Force extension relations of a Gaussian chain and a Langevin chain are compared in \fref{force_extension}, which clearly shows divergence at higher extensions resulting in a ``hardening" characteristic. Note that in the small extension limit ($\textsc{z}/(Na)\lesssim 1/3$), the force-extension curve for Langevin chain and Gaussian chain are indistinguishable. 

To find stress in a brush with Langevin chains, we extend SST-L described in \cite{biesheuvel2008}. Calculation of stress using \eqref{eq2p3} requires us to first determine free energy density $f$, and the derivatives $\frac{\partial f}{\partial \epsilon_{\textsc{xx}}}$ and $\frac{\partial \epsilon_{\textsc{zz}}}{\partial \epsilon_{\textsc{xx}}}$. In the following, we first calculate free energy density, followed by the derivative terms, and ultimately calculate stress variation in brushes of varying graft densities, numerically.

\subsubsection{Free energy density}
Free energy density in a brush with Langevin chains has two contributors: (a) Langevin chain elasticity, $f_{el}$ and, (b) interactions among the monomers, $f_{int}$. Intuitively, the area under force-extension curve in~\fref{force_extension} furnishes the free energy of elastic stretching of a single chain, $F_{chain}$, as:
\begin{align}
F_{chain}&=\int_0^{\textsc{z}}p^{\prime}(\textsc{z}^{\prime})d\textsc{z}^{\prime}=p\textsc{z}-\int_0^p\textsc{z}^{\prime}(p^{\prime})dp^{\prime} \nonumber \\
&=\left( \bar{p}\bar{\textsc{z}}-\log\left(\frac{\sinh(\bar{p})}{\bar{p}}\right)\right)k_BTN, \quad \bar{p}=\frac{pa}{k_BT}, \quad \bar{\textsc{z}}=\frac{\textsc{z}}{Na}. \label{eq2p8}
\end{align}
In the previous equation, the complementary energy (second term) is evaluated inside the integral first and then subtracted from the total energy (first term). Also notice that the stretching force and height have been normalized in the above.

A polymer chain in a brush is like a chain in a one dimensional external field. This results in a stretching force $\bar{p}$ in the chain which varies along the chain length. To find stretching free energy in this case, we consider chain segments in a slit of width $d\textsc{z}$ in the brush, as shown in~\fref{brush_slit}. Assuming that there are $dn$ monomers of a chain segment within this slit, the  free energy is obtained from~\eqref{eq2p8} by replacing $N$ with $dn$:
\begin{equation}
dF_{chain}=\frac{k_BT}{ae(\bar{p})}\left(\bar{p}e(\bar{p})-\log\left(\frac{\sinh(\bar{p})}{\bar{p}} \right)\right)d\textsc{z} , \quad e=\frac{1}{a}\frac{d\textsc{z}}{dn}=\mathcal{L}(\bar{p}). \label{eq2p9}
\end{equation}
Moreover, the stretching force ($\bar{p}$) in the above depends on height $\bar{\textsc{z}}$ of the segment above the grafting surface as well as the height of the chain end $\bar{\zeta}$ ($=\zeta/(Na)$). For a given mean potential field ($V(\textsc{z})=\bar{V}(\bar{\textsc{z}})$), $\bar{p}$ at height $\bar{\textsc{z}}$ for a chain with end at $\bar{\zeta}$ is obtained by the following relation~\cite{amoskov94}:
\begin{equation}
\log\left( \frac{\sinh(\bar{p})}{\bar{p}} \right)=\bar{V}(\bar{\zeta})-\bar{V}(\bar{\textsc{z}}), \label{eq2p10b}
\end{equation}
where $\bar{V}(\bar{\textsc{z}})$ is given by~\cite{biesheuvel2008}:
\begin{equation}
\bar{V}(\bar{\textsc{z}})=2\frac{\bar{\textsc{z}}^2}{a}\left(\frac{2-\frac{4}{5}\bar{\textsc{z}}^2}{1-\bar{\textsc{z}}^2}\right). \label{eq2p11}
\end{equation}
So, we can conclude that end density, $g(\zeta)$, plays a significant role in determining the stretching free energy density at $\textsc{z}$. Now, the total elastic free energy density of the brush at height $\textsc{z}$ can be written in terms of end-density distribution function, $g(\zeta)$, as:
\begin{align}
f_{el}=&\int_{\textsc{z}}^H\frac{dF_{chain}}{d\textsc{z}}g(\zeta)d{\zeta}=\int_{\bar{\textsc{z}}}^{\bar{H}}\frac{k_BT}{ae(\bar{p})}\left(\bar{p}e(\bar{p})-\log\left(\frac{\sinh(\bar{p})}{\bar{p}} \right)\right)\bar{g}(\bar{\zeta})d{\bar{\zeta}},  \nonumber \\
&\bar{H}=\frac{H}{Na}, \qquad \bar{g}(\bar{\zeta})=Nag(\zeta). \label{eq2p10}
\end{align}
Note that the above form of $\bar{V}(\bar{\textsc{z}})$ is an empirical rational fraction approximation to the accurate power series in $\bar{\textsc{z}}$ reported in \cite{amoskov94}. Also, calculation of $\bar{g}(\bar{\zeta})$ follows the description in \cite{biesheuvel2008} and is briefly summerized in \ref{SSTLcalc}. A (semi)analytical procedure to find $f_{el}$ at a given $\bar{\textsc{z}}$, then is to find $\bar{V}$ in~\eqref{eq2p11} first, followed by solving for $\bar{p}$ in~\eqref{eq2p10b} and $\bar{g}(\bar{\zeta})$ in \eqref{eq2p14}, and finally using $\bar{p}$ and $\bar{g}(\bar{\zeta})$ in~\eqref{eq2p10}.

To calculate $f_{int}$, we make use of CS equation of state for hard sphere mixtures with a correction for connection between monomers in a polymer chain~\cite{biesheuvel2008}. Here, polymer chains are viewed as a series of beads of volume $A_0 a$, where $A_0=\frac{\pi}{6}d^2$, $d$ is size of a bead and $a$ is length of a bead. Using the following relation between the chemical potential and free energy from~\cite{shim89},
\begin{equation}
\mu d=\frac{1}{k_BT}\frac{\partial f_{int}}{\partial \phi}=\frac{A_0 a}{k_BT}\frac{\partial f_{int}}{\partial V_f}, \label{eq2p11b}
\end{equation}
where $\phi$ is monomer density, and $V_f$ is volume fraction ($V_f=A_0 a \phi$), the interaction free energy density follows:
\begin{equation}
f_{int}=\frac{k_BT}{A_0 a}\int_0^{V_f}d\tilde{\mu}(V_f)dV_f, \label{eq2p12}
\end{equation}
where $\mu=\mu(\phi)=\tilde{\mu}(V_f)$. Chemical potential per unit chain length ($\tilde{\mu}(V_f)$) from the modified CS equation of state is~\cite{biesheuvel2008}:
\begin{equation}
\tilde{\mu}(V_f)=\frac{1}{d}\left( V_f\frac{7-7V_f+2V_f^2}{(1-V_f)^3}+\log(1-V_f)\right). \label{eq2p13}
\end{equation}
See \ref{SSTLcalc} for the calculation of $V_f$. In summary, to numerically calculate $f_{int}$, $V_f$ is calculated first for a brush of a given height (see \ref{SSTLcalc}), followed by \eqref{eq2p13} to find $\tilde{\mu}(V_f)$, which in turn is used in \eqref{eq2p12} to determine $f_{int}$.

\subsubsection{Calculation of the derivatives}
\label{deriv_calc}
In this section, we first calculate $\frac{\partial \epsilon_{\textsc{zz}}}{\partial \epsilon_{\textsc{xx}}}$ and $\frac{\partial f}{\partial \epsilon_{\textsc{xx}}}$, and subsequently numerically calculate the stress variation $\sigma_{\textsc{xx}}(\bar{\textsc{z}})$. The derivative of $\epsilon_{\textsc{zz}}$ with respect to the applied strain, $\epsilon_{\textsc{xx}}$, can be expressed as:
\begin{equation}
\frac{\partial \epsilon_{\textsc{zz}}}{\partial \epsilon_{\textsc{xx}}}=\frac{\partial}{\partial \epsilon_{\textsc{xx}}}\left(\frac{\partial \bar{u}}{\partial \bar{\textsc{z}}}\right)=\frac{\partial}{\partial \bar{\textsc{z}}}\left(\frac{\partial \bar{u}}{\partial \epsilon_{\textsc{xx}}}\right),
\label{eq2p19}
\end{equation}
where $\bar{u}=u/(Na)$ and $u$ is the displacement of a thin layer at $\bar{\textsc{z}}$ (see~\fref{brush_slit}) in the $\textsc{z}$-direction due to the applied strain. Finding the above derivative requires us to first find $\frac{\partial \bar{u}}{\partial \epsilon_{\textsc{xx}}}$. To this end, we make use of the fact that the number of monomers within a thin layer of volume $V_0$ at height $\bar{\textsc{z}}$ (see~\fref{brush_slit}), $\phi V_0$, does not change due to the applied strain ($\Delta(\phi V_0)=\Delta(V_f V_0)/(A_0 a)=0$), which yields~\cite{manav2018}:
\begin{equation}
\frac{\partial}{\partial \bar{\textsc{z}}}\left(\frac{\partial \bar{u}}{\partial \epsilon_{\textsc{xx}}}\right)=-\frac{1}{V_f}\frac{\partial V_f}{\partial \epsilon_{\textsc{xx}}}-1,
\label{eq2p21}
\end{equation}
with the boundary condition:
\begin{equation}
\left[\frac{\partial \bar{u}}{\partial \epsilon_{\textsc{xx}}} \right]_{\bar{\textsc{z}}=\bar{H}}=\frac{\partial \bar{H}}{\partial \epsilon_{\textsc{xx}}}.
\label{eq2p22}
\end{equation}
The derivative of volume fraction with respect to the applied strain, $\frac{\partial V_f}{\partial \epsilon_{\textsc{xx}}}$, is obtained by taking derivative of both sides in \eqref{eq2p0}.
\begin{align}
\frac{\partial V_f}{\partial \epsilon_{\textsc{xx}}}=&\frac{8}{5}\frac{d}{a}\left(\frac{\bar{H}(2\bar{H}^4-4\bar{H}^2+5)}{(1-\bar{H}^2)^2}\frac{\partial \bar{H}}{\partial \epsilon_{\textsc{xx}}}- \frac{\bar{\textsc{z}}(2\bar{\textsc{z}}^4-4\bar{\textsc{z}}^2+5)}{(1-\bar{\textsc{z}}^2)^2}\frac{\partial \bar{u}}{\partial \epsilon_{\textsc{xx}}} \right) \nonumber \\
& \times \frac{(1-V_f)^4}{6+3V_f-4V_f^2+V_f^3}.
\label{eq2p20}
\end{align}
The derivative of $\bar{H}$ with respect to the applied strain is evaluated numerically.
\begin{equation}
\frac{\partial \bar{H}}{\partial \epsilon_{\textsc{xx}}}=\frac{\partial \rho_g}{\partial \epsilon_{\textsc{xx}}}\frac{\partial \bar{H}}{\partial \rho_g}=-\rho_g\frac{\partial \bar{H}}{\partial \rho_g},
\label{eq2p18}
\end{equation}
and $\frac{\partial \bar{H}}{\partial \rho_g}$ is obtained by finding $\bar{H}=\bar{H}(\rho_g)$ using \eqref{eq2p13b}.

\begin{figure}[!h]
\center
\includegraphics[width=8.7cm]{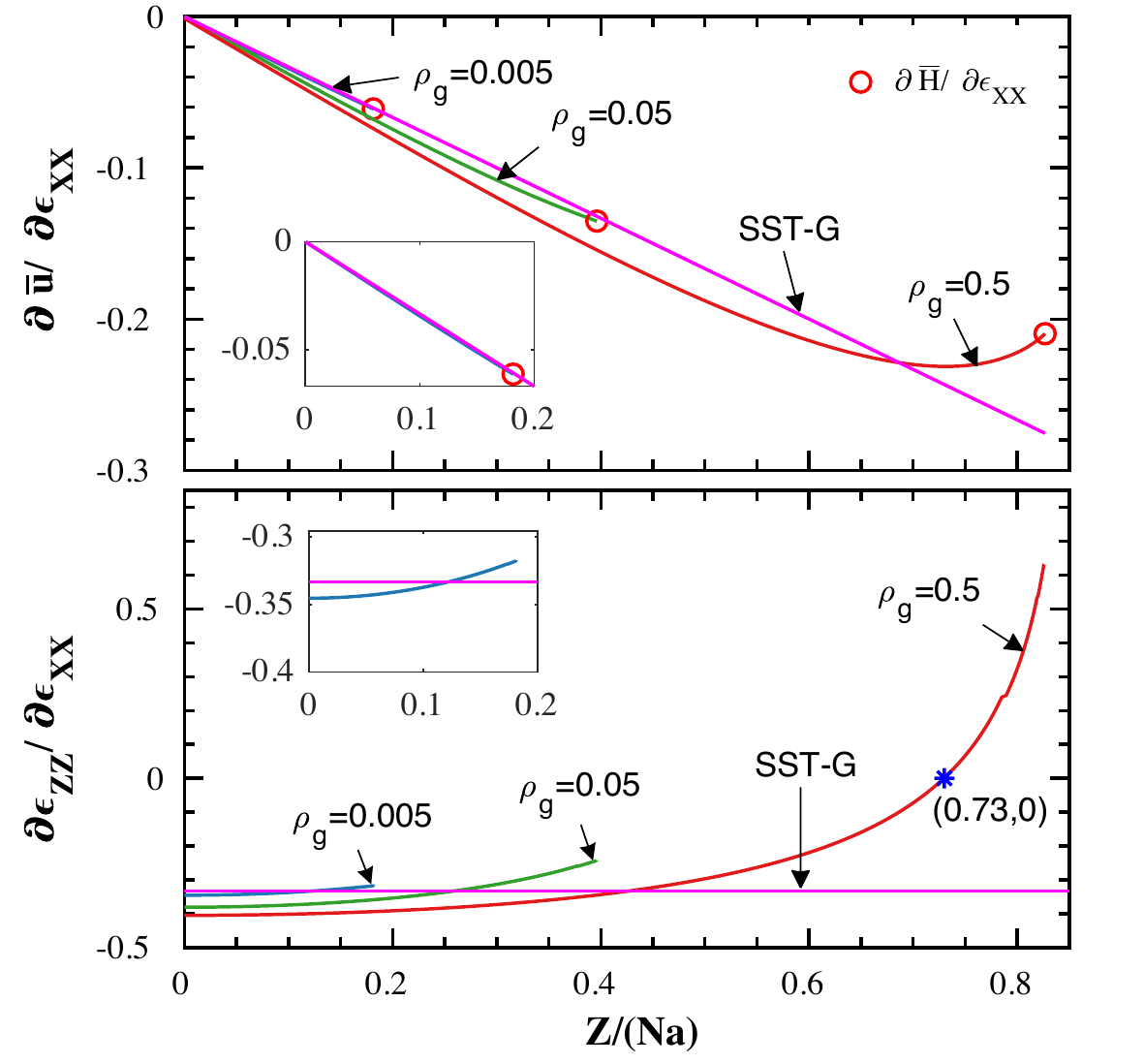}
\caption{The figure shows $\frac{\partial \bar{u}}{\partial \epsilon_{\textsc{xx}}}$ and $\frac{\partial \epsilon_{\textsc{zz}}}{\partial \epsilon_{\textsc{xx}}}$ vs $\bar{\textsc{z}}$ curves for three graft densities along with $\frac{\partial \bar{H}}{\partial \epsilon_{\textsc{xx}}}$. Analytical relations obtained from SST-G are also plotted. As $\frac{\partial \epsilon_{\textsc{zz}}}{\partial \epsilon_{\textsc{xx}}}$ is obtained by numerical differentiation of $\frac{\partial \bar{u}}{\partial \epsilon_{\textsc{xx}}}$ (see \eqref{eq2p19}), there are jumps at the end of curves in the lower plot but they are not shown. In the inset, numerical curve for the lowest graft density is compared with the analytical curve and a close agreement is observed, specifically in $\frac{\partial \bar{u}}{\partial \epsilon_{\textsc{xx}}}$ plot. However, considerable deviation is observed for high graft densities. Also, for very high graft density, $\frac{\partial \epsilon_{\textsc{zz}}}{\partial \epsilon_{\textsc{xx}}}$ becomes positive for $\bar{\textsc{z}}>0.73$, which is not captured by SST-G.}
\label{poisson_like_ratio}
\end{figure}

After substituting \eqref{eq2p20} in \eqref{eq2p21}, \eqref{eq2p21} with the boundary condition \eqref{eq2p22} is solved numerically to obtain $\frac{\partial \bar{u}}{\partial \epsilon_{\textsc{xx}}}$ and subsequently $\frac{\partial V_f}{\partial \epsilon_{\textsc{xx}}}$. By taking derivative of $\frac{\partial \bar{u}}{\partial \epsilon_{\textsc{xx}}}$ with respect to $\bar{\textsc{z}}$, we obtain $\frac{\partial \epsilon_{\textsc{zz}}}{\partial \epsilon_{\textsc{xx}}}$.   \fref{poisson_like_ratio}  shows the variation of $\frac{\partial \bar{u}}{\partial \epsilon_{\textsc{xx}}}$ and $\frac{\partial \epsilon_{\textsc{zz}}}{\partial \epsilon_{\textsc{xx}}}$ with $\bar{\textsc{z}}$ for three graft densities and compares the numerically obtained curves from SST-L with analytical relation $\frac{\partial \bar{u}}{\partial \epsilon_{\textsc{xx}}}=-\frac{1}{3}\bar{\textsc{z}}$ (and $\frac{\partial \epsilon_{\textsc{zz}}}{\partial \epsilon_{\textsc{xx}}}=-\frac{1}{3}$) obtained from SST-G \cite{manav2018}. For the lowest graft density, curves obtained from SST-L and SST-G agree well, as expected. This can be seen in the insets of  \fref{poisson_like_ratio} where the values of $\frac{\partial \bar{u}}{\partial \epsilon_{\textsc{xx}}}$ are almost on top of the SST-G prediction while for $\frac{\partial \epsilon_{\textsc{zz}}}{\partial \epsilon_{\textsc{xx}}}$ only small deviations are seen. However, as graft density is increased, deviations from the SST-G theory become more apparent as depicted in the plot for $\rho_g=0.05$ and 0.5.

We also observed another interesting feature predicted by the SST-L. For $\rho_g \ge 0.5$,  the predicted values of $\frac{\partial \epsilon_{\textsc{zz}}}{\partial \epsilon_{\textsc{xx}}}$ for $\bar{\textsc{z}} > 0.73$ become positive indicating that the layers above this height undergo an expansion when the brush is stretched in $\textsc{x}$-direction. The critical point at which this occurs, is highlighted in the plot with a blue marker (*). This change in the sign of strain is observed only for very high graft density brushes and is only captured by SST-L. Note that, to smooth the curve for $\frac{\partial \epsilon_{\textsc{zz}}}{\partial \epsilon_{\textsc{xx}}}$ obtained by numerical differentiation of $\frac{\partial \bar{u}}{\partial \epsilon_{\textsc{xx}}}$, a high order polynomial was fit to $\frac{\partial \bar{u}}{\partial \epsilon_{\textsc{xx}}}$ vs $\bar{\textsc{z}}$ curve and the fitted polynomial was differentiated.

To evaluate derivative of free energy density at height $\bar{\textsc{z}}$ with respect to the applied strain, we find the derivative of interaction part, $\frac{\partial f_{int}}{\partial \epsilon_{\textsc{xx}}}$, and stretching part, $\frac{\partial f_{el}}{\partial \epsilon_{\textsc{xx}}}$, independently, and then sum them up. $\frac{\partial f_{int}}{\partial \epsilon_{\textsc{xx}}}$ is obtained by taking derivative of \eqref{eq2p12}:
\begin{equation}
\frac{\partial f_{int}}{\partial \epsilon_{\textsc{xx}}}=\frac{k_BT}{A_0 a}\left( V_f\frac{7-7V_f+2V_f^2}{(1-V_f)^3}+\log(1-V_f) \right)\frac{\partial V_f}{\partial \epsilon_{\textsc{xx}}}. \label{eq2p23}
\end{equation}

Finding $\frac{\partial f_{el}}{\partial \epsilon_{\textsc{xx}}}$ is more involved. Taking derivative of the expression for $f_{el}$ in \eqref{eq2p10} gives:
\begin{align}
\frac{\partial f_{el}}{\partial \epsilon_{\textsc{xx}}}=&\frac{k_BT}{a}\int_{\bar{\textsc{z}}}^{\bar{H}}\log\left(\frac{\sinh(\bar{p})}{\bar{p}} \right)\left(\frac{1-(\coth(\bar{p}))^2+\frac{1}{\bar{p}^2}}{(e(\bar{p}))^2}\right)\bar{g}(\bar{\zeta})\frac{\partial \bar{p}}{\partial \epsilon_{\textsc{xx}}}d\bar{\zeta} \nonumber \\
&+\frac{k_BT}{a}\int_{\bar{\textsc{z}}}^{\bar{H}} \left(\bar{p}-\frac{1}{e(\bar{p})}\log\left(\frac{\sinh(\bar{p})}{\bar{p}} \right)\right) \frac{\partial \bar{g}(\bar{\zeta})}{\partial \epsilon_{\textsc{xx}}}d\bar{\zeta}, \label{eq2p24}
\end{align}
To evaluate the above relation, we need to obtain the derivative of local stretching force ($\frac{\partial \bar{p}}{\partial \epsilon_{\textsc{xx}}}$) and normalized end density ($\frac{\partial \bar{g}(\bar{\zeta})}{\partial \epsilon_{\textsc{xx}}}$). By making use of the implicit relation involving $\bar{p}$ in \eqref{eq2p10b} and recognizing that $\frac{\partial \bar{V}(\bar{\zeta})}{\partial \epsilon_{\textsc{xx}}}=0$ as $\bar{\zeta}$ is the integration variable in \eqref{eq2p24}, we obtain the desired derivative:
\begin{equation}
\frac{\partial \bar{p}}{\partial \epsilon_{\textsc{xx}}}=-\frac{1}{e(\bar{p})}\frac{\partial \bar{V}(\bar{\textsc{z}})}{\partial \epsilon_{\textsc{xx}}}=-\frac{1}{e(\bar{p})}\frac{8}{5}\frac{\bar{\textsc{z}}}{a}\frac{2\bar{\textsc{z}}^4-4\bar{\textsc{z}}^2+5}{(1-\bar{\textsc{z}}^2)^2}\frac{\partial \bar{u}}{\partial \epsilon_{\textsc{xx}}}. \label{eq2p25}
\end{equation}

\begin{figure}[!h]
\center
\includegraphics[width=8.7cm]{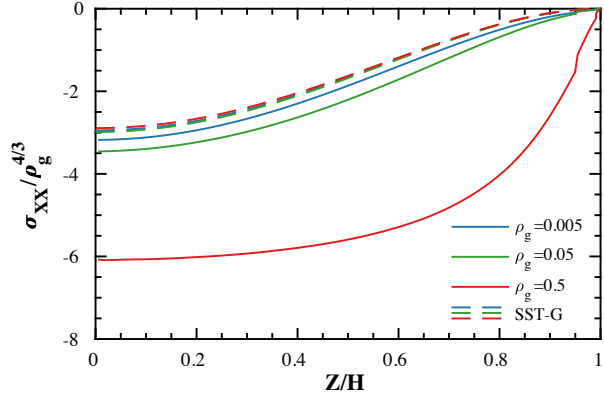}
\caption{Stress profiles obtained from SST-L is compared with the predictions from SST-G. Curves obtained from SST-G for different graft densities fall on top of each other. SST-G and SST-L curves corresponding to the lowest graft density are very close. However, at high graft density, SST-L predicts much higher stress. The jump in stress profile near the top of a brush is a numerical \emph{artifact}.}
\label{stress_Z_comparison}
\end{figure}

See \ref{SSTLcalc} for calculation of $\frac{\partial \bar{g}(\bar{\zeta})}{\partial \epsilon_{\textsc{xx}}}$. On solving the above numerically and substituting the values of $f$, $\frac{\partial f}{\partial \epsilon_{\textsc{xx}}}$ and $\frac{\partial \epsilon_{\textsc{zz}}}{\partial \epsilon_{\textsc{xx}}}$ in~\eqref{eq2p3}, we obtain the stress profile $\sigma_{\textsc{xx}}(\bar{\textsc{z}})$. SST-G and SST-L stress profiles are compared  in \fref{stress_Z_comparison}. Note that for an accurate comparison between the two theories, we prescribe the same brush height for SST-G as given by SST-L for a given graft density. This allows determination of excluded volume parameter $v$ in SST-G for each graft density, and subsequent calculation of $\sigma_{\textsc{xx}}$ using \eqref{eq2p6} (See \ref{SSTLcalc} for monomer density and end density comparisons). Based on \eqref{eq2p6}, we expect $\sigma_{\textsc{xx}}/\rho_g^{4/3}$ vs $\textsc{z}/H$ curves for different graft densities obtained from SST-G to fall on a master curve. The small deviations observed in \fref{stress_Z_comparison} are due to a very small difference in excluded volume parameters for different graft densities. Turning our attention to the values predicted by SST-L, we observe that for small graft densities the predictions are close to SST-G. However, for large values of graft density, we observe that the prediction of stress distribution changes significantly with changes in the shape of the distribution. Note that the jump in stress profile near the top of a brush in \fref{stress_Z_comparison} is a numerical artifact and occurs due to the fact that the end density shows sharp descent near the top of the brush (see~\fref{endden_Z_comparison}) and numerical evaluation of the derivative of the end density (in \eqref{eq2p26}) near the top requires much smaller step size than the step size in the rest of the brush. The jump is observed at the $\bar{\textsc{z}}$ where step size changes.

%%%%%%%%%%%%%%%%%%%%%%%%%%%%%%%%%%%%%%%%%%%%%%

\section{Molecular dynamics simulation}
\label{MD}
The purpose of MD simulations is to verify the predictions of SST-G and SST-L without placing any restrictions \emph{a priori} on (a) virial truncation, (b) Langevin or Gaussian assumptions for chain elasticity, and (c) classical paths restriction on chain conformations. 

\begin{figure}[!h]
\center
\includegraphics[width=8.7cm]{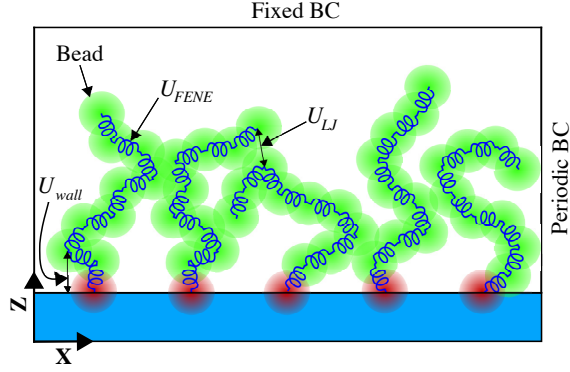}\\
\caption{Front view of MD simulation box with bead spring chains representing polymers, along with the interactions involved in the brush and the boundary conditions. Red coloured beads are fixed to the rigid substrate.}
\label{MD_intro}
\end{figure}

We use the Large-scale Atomic/Molecular Massively Parallel Simulator (LAMMPS)~\cite{plimpton95} code, to simulate a neutral polymer brush grafted to a rigid substrate. A cartoon of our model is illustrated in \fref{MD_intro}. Let us now consider a system of $N_g$ chains with each chain made of $N+1$ beads. The first bead of each chain is fixed to the substrate. The total number of unconstrained beads in the system is $N_{tot}=N_g N$. Here, we perform a Langevin dynamics simulation wherein temperature is controlled by attaching a heat bath to each of the unconstrained beads. Consequent coupling results in a random force on each bead along with a viscous force governed by the fluctuation-dissipation theorem. The governing Langevin stochastic differential equation to be solved then is:
\begin{equation}
\centering
m^i\frac{{\rm d}^2{\bold r}^i}{{\rm d}t^2}=-\frac{\partial U}{\partial \bold{r}^i}-\Gamma\frac{{\rm d}{\bold r}^i}{{\rm d}t}+\bold{F}^i(t),\quad i=1\dots N_{tot},
\label{eq4p1}
\end{equation}
where $m^i$ and $\bold{r}^i$ are mass and position, respectively, of the $i^{\rm th}$ unconstrained bead, $U$ is the total potential energy of the system, and $\Gamma$ is bead friction. In the simulations, $\Gamma=2.0\tau^{-1}$, where $\tau$ is unit of time in Lennard-Jones (LJ) units. Note that LJ units are used throughout the MD simulation section. $\bold{F}^i(t)$ is a Gaussian white noise satisfying the following relation:
\begin{equation}
\left<\bold{F}^i(t)\cdot\bold{F}^j(t')\right>=\delta_{ij}\delta(t-t')6k_BT\Gamma,\quad i,~j=1\dots N_{tot},
\label{eq4p2}
\end{equation}
in accordance with the fluctuation-dissipation theorem~\citep{kubo}. Note that $\delta_{ij}$ is Kronecker delta function and $\delta(\cdot)$ is Dirac delta function.

\begin{figure}[!h]
\center
\includegraphics[width=8.7cm]{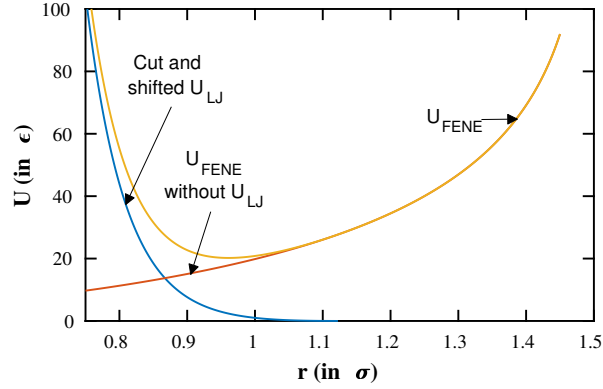}
\caption{Variation of the pair ($U_{LJ}$ cut at its minimum and shifted up) and bond ($U_{FENE}$) potentials with the distance between interacting beads. Observe the short range repulsion in $U_{FENE}$ is due to the LJ term present in it (see \eqref{eq4p4}).}
\label{MD_intro-b}
\end{figure}

The total potential energy of the system has three-main contributions: (i) bond potential $U_{\text{FENE}}$, (ii) non-bonded pair potential $U_{LJ}$, and (iii) potential governing interaction of the beads with the grafting surface $U_{wall}$. 
\begin{align}
\centering
U=& \frac{1}{2}\sum_{i=1}^{\substack{N_{tot}}}\left(\sum_{j=1}^{N_{tot}}U_{\text{FENE}}(r^{ij})+\sum_{k=1}^{N_g}U_{\text{FENE}}(r_g^{ik})\right)+\nonumber\\ 
&\frac{1}{2}\sum_{i=1}^{N_{tot}}  \sum_{\substack{j=1\\ j\ne i\pm1}}^{N_{tot}}U_{LJ}(r^{ij})+\sum_{i=1}^{N_{tot}}U_{wall}(\textsc{z}^i),
\label{eq4p3}
\end{align}
where $r^{ij}=|\bold{r}^{i}-\bold{r}^{j}|$ is the distance between beads $i$ and $j$, and $r_g^{ik}=|\bold{r}^{i}-\bold{r_g}^{k}|$ is the distance between bead $i$ and constrained (grafted) bead $k$. $\textsc{z}^i$ is the perpendicular distance of a bead from the grafting surface. Also, the beads fixed to a substrate interact only with the unconstrained bead bonded to it. Beads of unit mass are connected by finite extensible nonlinear elastic (FENE) spring representing a bond between effective monomers as done in earlier MD studies on brushes \cite{murat89,grest93}. The potential associated with FENE springs is given as (see \fref{MD_intro-b}):
\begin{equation}
U_{\text{FENE}}(r^{ij})=b^{ij}\left[-0.5KR_0^2\log\left(1-\left(\frac{r^{ij}}{R_0}\right)^2\right)+4\epsilon p_c\left(\left(\frac{\sigma}{r^{ij}}\right)^{12}-\left(\frac{\sigma}{r^{ij}}\right)^6+\frac{1}{4}\right)\right],
\label{eq4p4}
\end{equation}
where  $b^{ij}$ is a bond order parameter that is 1 for adjacent beads in a polymer chain, and 0 otherwise, $K$ is a constant determining stiffness, $R_0$ is the maximum extension in the spring. $\epsilon$ and $\sigma$ are the energy and length scales associated with the second term which is LJ potential. $p_c$ is a piecewise continuous function used to truncate the LJ potential to only account for repulsion forces. Thus, $p_c=1$ for $r^{ij}\le r_c=2^{1/6}\sigma$ and is $0$ otherwise. Note that this $\sigma$ is different from the symbol for stress tensor ($\sigma_{\textsc{ij}}$), which always has a subscript in this work. 

The first term in the expression above is attractive and is balanced by the repulsive second term at equilibrium bond length. In the simulation, $R_0=1.5\sigma$. In an athermal simulation at reduced temperature $T=1.2\epsilon/k_B$ and for $K=30\epsilon/\sigma^2$, average bond length is equal to $0.97\sigma$. So, while comparing MD simulation results with SST-L, we take $a=0.97 \sigma$ and $d=\sigma$ in interpreting MD results. \fref{MD_intro-b} shows the two terms of the FENE potential, and the total interaction potential as previously described.

The interaction between nonbonded beads is governed by LJ potential with appropriate cut-off (see \fref{MD_intro-b}). 
\begin{align}
U_{LJ}(r^{ij})=
\begin{cases}
4\epsilon\left(\left(\frac{\sigma}{r^{ij}}\right)^{12}-\left(\frac{\sigma}{r^{ij}}\right)^6-\left(\frac{\sigma}{r_c}\right)^{12}+\left(\frac{\sigma}{r_c}\right)^6\right) & r^{ij}\le r_c,\\
0 & r^{ij}>r_c,
\end{cases}
\label{eq4p5}
\end{align}
where $r^{ij}$ is the distance between a pair of interacting monomers, and $r_c$ is the cut-off distance. To simulate brush in a good solvent condition, $r_c=2^{1/6}\sigma$ such that pair interaction is purely repulsive. This is often referred to as athermal simulation since the potential is close to a hard sphere potential \cite{grest93}.

The polymer chains have their one end fixed to a rigid wall. To ensure that the polymer chains do not cross the wall, the bead-wall interaction is repulsive, and governed by the following potential:
\begin{align}
U_{wall}(\textsc{z}^i)=
\begin{cases}
4\epsilon\left(\left(\frac{\sigma}{\textsc{z}^i}\right)^{12}-\left(\frac{\sigma}{\textsc{z}^i}\right)^6-\left(\frac{\sigma}{\textsc{z}_c}\right)^{12}+\left(\frac{\sigma}{\textsc{z}_c}\right)^6\right) & \textsc{z}^i\le \textsc{z}_c=2^{1/6}\sigma,\\
0 & \textsc{z}^i>\textsc{z}_c=2^{1/6}\sigma.
\end{cases}
\label{eq4p6}
\end{align}

Athermal simulations at $T=1.2\epsilon/k_B$ are performed. Length and width of the simulation box and hence, of the grafting surface, is chosen to be the same and slightly larger than the brush height except for $\rho_g\ge0.2$, wherein to limit the total number of beads ($N_{tot}$) at $\sim 500,000$, the box size was smaller than the brush height. The first monomer of each of the chains is fixed to one of the uniformly spaced grid points on the grafting surface. A random walk conformation of a chain starting at each of the grafting points is obtained and used as the starting brush configuration. An efficient way to generate initial configuration, particularly for low graft graft density brushes, is described in \ref{init_config}. In the directions along the length and width of the box ($\textsc{x}$ and $\textsc{y}$), periodic boundary conditions are specified, see~\fref{MD_intro}. In $\textsc{z}$-direction, fixed boundary is specified and height of the simulation box is chosen sufficiently large so that no particle goes out of the box during a simulation. 

Particle velocities are randomly assigned to ensure a reduced temperature of $T=1.2\epsilon/k_B$. Note that the initial brush configuration may have an overlap between monomers. Because LJ potential is unstable when the distance between interacting particles approaches zero, we initially run the system with the following soft pair potential instead of LJ pair potential for $\sim 30,000$ time steps, before switching to the LJ potential.
\begin{align}
U_{soft}(r^{ij})=
\begin{cases}
A\left(1+\cos\left(\frac{\pi r^{ij}}{r_c} \right) \right) & r^{ij}\le r_c,\\
0 & r^{ij}>r_c,
\end{cases}
\label{eq4p7}
\end{align}
where $r^{ij}$ is the distance between a pair of interacting monomers. $r_c$ was chosen to be $\sigma$, and $A$ was increased from $0$ to $30\epsilon$ in 10 steps to ensure that the configuration becomes stable upon switching to LJ potential. After switching to LJ pair potential, the system is run for $\sim 10^7$ steps to equilibrate. Once the monomer density profile becomes stable, we run the system for another $\sim 5\times 10^6$ steps to obtain data to calculate property values. Note however that for very small graft densities, where length of each chain ($Na$) is large, equilibration took $\sim 10$ times more steps.

To obtain the variation of the brush properties, for example number density, end density, stress etc., with distance from the grafting surface, we divide the simulation volume in bins of thickness $\sigma$, and length and width along the grafting surface the same as that of the simulation box. Value of any of the above properties at the center of a bin is calculated by averaging the property values over the bin, and over the length of the simulation.

\subsection{Calculation of stress} \label{StressMD}
We take virial stress as the stress measure. At each time step, we compute the following quantity for the $i^{\text{th}}$ bead:
\begin{equation} \label{stress_per_atom}
S_{ab}^i = -\left[ m^i v^i_a v^i_b + \frac{1}{2} \sum_{j=1}^{N_p} \left( r^i_{a} P^{ij}_{b} + r^j_{a} P^{ji}_{b}  \right)+ \frac{1}{2} \sum_{k=1}^{N_b} \left( r^i_{a} p^{ik}_{b} + r^k_{a} p^{ki}_{b} \right)   \right]
\end{equation}
where $m^i$ is the mass of the $i^{\text{th}}$ bead, $r^i_{a}$ and $v_a^i$ are the $a^{\text{th}}$ component of position vector and velocity of the $i^{\text{th}}$ bead, $N_p$ and $N_b$ are the number of pair neighbors and bonds of the  $i^{\text{th}}$ bead, respectively. $P^{ij}_{b}$ is $b^{\text{th}}$ component of force on the $i^{\text{th}}$ bead due to pair interaction with the $j^{\text{th}}$ bead, and $p^{ik}_{b}$ is $b^{\text{th}}$ component of force on the $i^{\text{th}}$ bead due to bond interaction with the $k^{\text{th}}$ bead. Now, let us consider the $n^{\text{th}}$ bin with volume $V_{bin}$. It has $N_{bin}$ beads at the $l^{\text{th}}$ time step. The \emph{instantaneous} virial stress in the $n^{\rm th}$ bin, defined at the $l^{\text{th}}$ time step, is:
\begin{equation} \label{stress_per_bin}
[\sigma_{ab}^{n}]_{l} = \frac{\displaystyle\sum_{i=1}^{N_{bin}} S_{ab}^i}{V_{bin}},
\end{equation}
which accounts for the behavior of multiple beads in the bin. In order to report statistically meaningful quantities, we computed the averaged stress tensor per bin at the $k^{\text{th}}$ time step as follows:
\begin{equation} \label{stress_per_bin_averaged}
\langle \sigma_{ab}^{n} \rangle_k  =\frac{\displaystyle\sum_{i=1}^{N_k} [\sigma_{ab}^{n}]_{(k+i)}}{N_k},
\end{equation}
where $N_k$ is a number of consecutive time steps of the simulation. We observe that the number of time steps used to average the stress components needs to be large enough to reduce fluctuations and spurious measures that could appear during entropic oscillations of the polymer brush. Our simulations showed that for $N_k \ge 1000$ the results are insensitive to the choice of $N_k$. Thus, we took $N_k = 1000$. We systematically do this for multiple time instances ($N_I\sim 5000$) and report the phase-averaged virial stress components in the $n^{\text{th}}$ bin as:
\begin{equation} \label{stress_per_bin_phase_averaged}
\overline{ \sigma}_{ab}^{n}  = \frac{ \displaystyle\sum_{k=1}^{N_I} \langle \sigma_{ab}^{n} \rangle_k}{N_I}.
\end{equation}

We remark that the reported values of the components of virial stress reflect the stress state of a collection of beads, and not a point wise measure of stress in the system. To obtain a point wise measure, other stress metric \cite{Admal:2010} should be employed.
%%%%%%%%%%%%%%%%%%%%%%%%%%%%%%%%%%%%%%%%%%%%%%

\section{Results: MD vs. SST-G and SST-L} \label{result}
We compare the results from SST calculation and MD simulation in this section. Since we have used generic potentials in MD simulation, the results are qualitative and quantitative mapping to a physical system requires one to determine $\epsilon$ and $\sigma$ for the system first.

\begin{figure}[!h]
	\center
	\includegraphics[width=8.7cm]{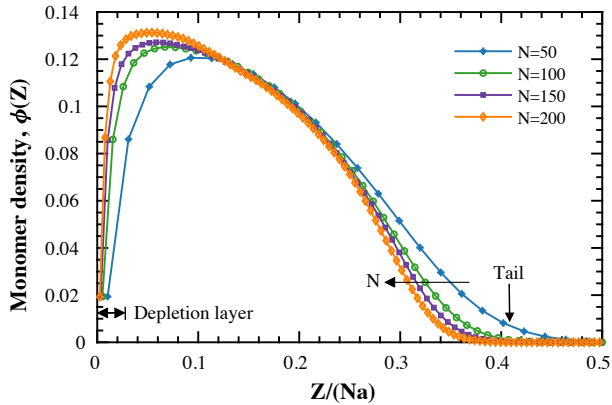}
	\caption{The effect of number of monomers in a chain ($N$) on the variation of monomer density with distance from the grafting surface in a polymer brush of graft density $0.03$. Monomer density curves converge to a single curve as $N$ is increased. Stretching parameter $\beta_s=14.66,~23.43,~32.29$ and $40.98$ for $N=50,~100,~150$ and $200$, respectively. From this, $N$ for each graft density is chosen such that $\beta_s>30$ in \emph{all} the athermal simulations carried out in this work. }
	\label{convergence_gd0p03}
\end{figure}

We studied brushes with graft densities ranging from $0.005$ to $0.5$. To start with, we ensure that chains are strongly stretched so that brush height is proportional to number of beads in a chain $N$. This allows an accurate comparison between MD and SST. To achieve this, we performed a convergence test wherein $N$ was increased to ensure that $\phi(\textsc{z})$ vs $\textsc{z}/(Na)$ curves for different $N$ converge to a single curve as shown in \fref{convergence_gd0p03}. We notice that the curves converge towards a single curve as $N$ is increased. Also notice the depletion layer and tail in the monomer density profile which are not present in SST predictions. They naturally appear in simulations, and shrink with an increasing $N$ as expected from numerical SCFT \cite{netz98}. Guided by this convergence test, we specify a minimum stretching parameter $\beta_s=3/2(H^2/(Na^2))>30$ as the convergence criterion for all the graft densities simulated. Note that it is shown in \cite{netz98} that for large $\beta_s$, numerical SCFT monomer density profiles agree well with SST profiles. We make use of this result by choosing $N$ according to this criterion. This ensures that polymer chains in a brush are stretched to a size at least $\gtrapprox 4.5$ times the end to end distance of corresponding ideal chain with no interacting chains nearby. We could not choose a higher threshold for $\beta_s$ because that would have required an exceedingly large $N$ for low graft densities, incurring much higher computational cost to reach equilibration.

\subsection{Monomer density}
\label{Numden_result}
\begin{figure}[!h]
	\center
	\includegraphics[width=8.6cm]{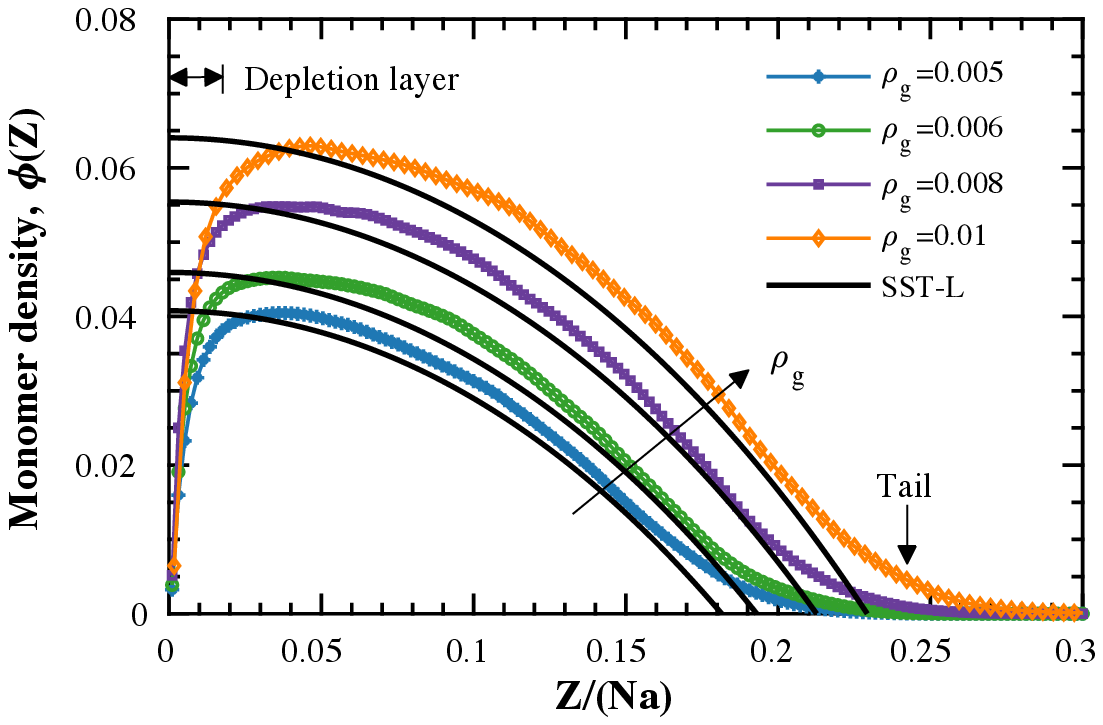}\\
	\includegraphics[width=8.6cm]{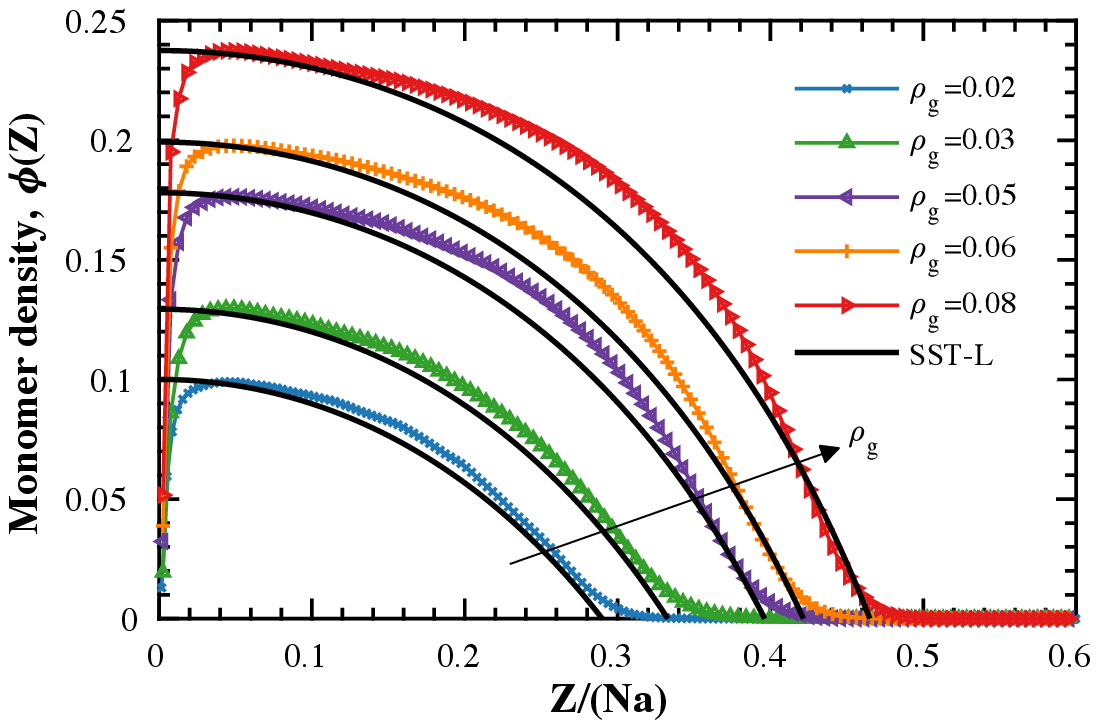} \\
	\includegraphics[width=8.6cm]{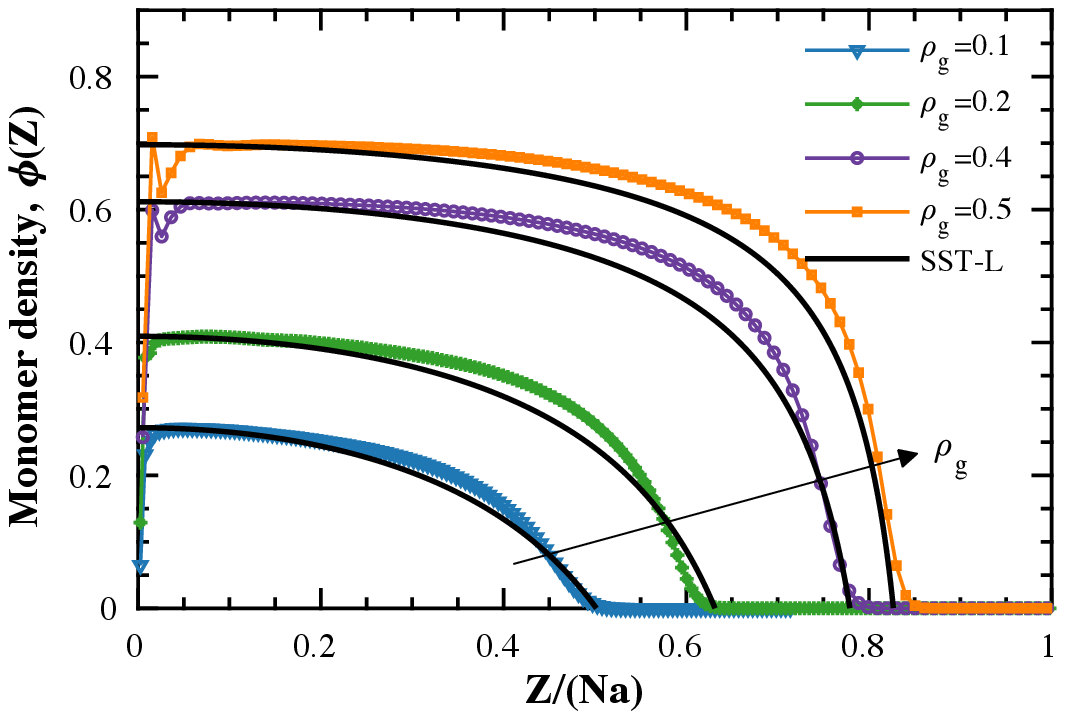}
	\caption{Variation of monomer density in a polymer brush with the distance from the grafting surface. Parabolic profile at low graft density (top) smoothly transitions to a step-like profile with increasing graft density (bottom). Note that density profiles obtained from SST-L agree well with MD predictions without the need for a fitting parameter.}
	\label{monomer_density_Z}
\end{figure}

Monomer density in a brush varies with distance from the grafting surface as shown in \fref{monomer_density_Z}. We divided the range of graft densities simulated into three regimes: low graft density ($\rho_g<0.02$), intermediate graft density $0.02\le\rho_g<0.1$, and high graft density ($\rho_g\ge0.1$), and show different plots accordingly. Predictions from SST-G are expected to be valid \emph{only} in low graft density regime. Notice that SST-L \cite{biesheuvel2008} closely approximates the monomer density profile for \emph{all} graft densities simulated, and the agreement improves with increasing graft density. Interestingly, the simulations naturally predict a smooth transition from a parabolic profile to a step like profile as graft density is increased.

\begin{figure}[!h]
	\center
	\includegraphics[width=8.7cm]{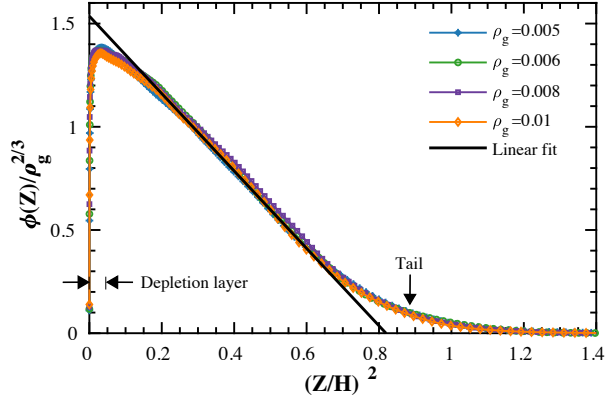}
	\caption{SST-G predicts the $\phi(\textsc{z})/\rho_g^{2/3}$ vs $(\textsc{z}/H)^2$ plot to be independent of $\rho_g$ and a straight line. The simulation shows good match with the theory in the bulk of the brush for small graft densities.}
	\label{monomer_density_Zparam}
\end{figure}

To validate monomer density profile predicted by SST-G in the low graft density regime, we plot scaled monomer density ($\phi(\textsc{z})/\rho_g^{2/3}$) with scaled distance from the grafting surface $((\textsc{z}/H)^2)$ in \fref{monomer_density_Zparam}. This plot validates two predictions, first, that $\phi(\textsc{z})\sim\rho_g^{2/3}$, and second, that the the monomer density has a parabolic profile (shows quadratic variation with distance from the grafting surface). This is clearly highlighted in the plot, where MD points fall on a line in the middle region of the polymer brush. However, we notice that the profile deviates from a parabola at the grafted as well as free end due to the effect of depletion layer and tail. The profile increasingly deviates from these predictions as graft density is increased.

Due to the presence of a tail in monomer density profile, height is difficult to identify clearly. So, average height is defined as the first moment of monomer density \cite{murat89}:
\begin{equation}
H:=\frac{8}{3}\frac{\int_0^{\infty}\textsc{z}\phi(\textsc{z})d\textsc{z}}{\int_0^{\infty}\phi(\textsc{z})d\textsc{z}}.
\label{eq3p9}
\end{equation}
The normalizing pre-factor $8/3$ ensures  that the height predicted by SST-G matches with the calculation above if parabolic monomer density profile obtained from SST-G is used in the above formula.

\begin{figure}[!htb]
	\center
	\includegraphics[width=8.7cm]{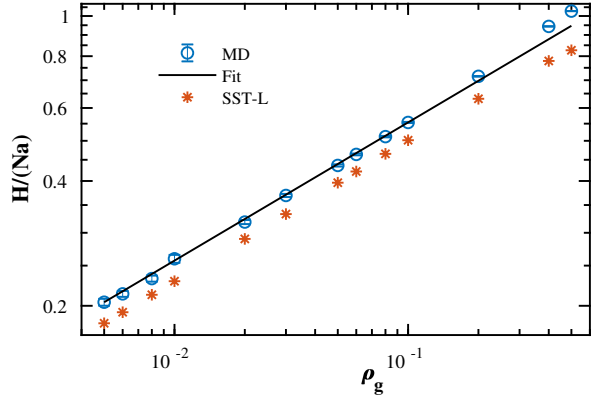}
	\caption{Variation of the height of a brush with graft density. A line of slope $0.33$ is drawn to show the match between MD prediction and SST-G and scaling theory. Height obtained from SST-L is also shown. Notice the deviation from linear fit at high graft densities.} 
	\label{H_rho_g}
\end{figure}

The dependence of brush height on graft density is shown in \fref{H_rho_g}. For the graft densities studied in this work, the scaling of height with respect to graft density matches closely with the theoretical prediction for $\rho_g\le0.2$. However, a deviation can be observed on increasing graft density further. The plot also shows the height predicted by SST-L which shows close agreement with the MD values. Interestingly, we observe an increase in slope of a curve obtained by joining MD points in~\fref{H_rho_g}, pointing to an increase in scaling exponent of $\rho_g$ in the expression for brush height from theoretically predicted $1/3$. However, height obtained from SST-L shows exactly the opposite trend. This discrepancy is an artifact of the way the height is calculated in \eqref{eq3p9}. Note that for a step profile without a depletion layer or tail at the ends of a brush, \eqref{eq3p9} predicts a height greater than the actual height of the brush. An unwanted consequence of this is that height predicted may be higher than the contour length of a chain ($H/(Na)>1$), as observed for the last point in \fref{H_rho_g} corresponding to $\rho_g=0.5$.

We use the the height obtained from MD to calculate $\beta_s$ and $\beta_e$. Table~\ref{beta_stretch} lists $N$, $\beta_s$, and $\beta_e$ for different graft densities. Note that $\beta_s>30$ for all graft densities. Extension in a chain $\beta_e$ helps determine the validity of Gaussian chain assumption. Based on $\beta_e$ values in Table~\ref{beta_stretch}, Gaussian elasticity is not valid for $\rho_g> 0.02$. To determine its validity at $\rho_g=0.02$, we need to consider nonuniform chain extension predicted by SST. Hence, we check the value of local stretching ($E(\textsc{z},\zeta)$) to determine validity of Gaussian assumption. As monomer density is highest close to the grafting surface, we find local stretching at $\textsc{z}\approx 0$ using \eqref{eq2p5d} from SST-G.
\begin{equation}
E(0,\zeta)=\frac{\pi \zeta}{2N}=\frac{\pi a}{2}\bar{\zeta}.
\label{eq3p10}
\end{equation}
$E(0,\zeta)$ is lower than $0.33$ for $\bar{\zeta}<\bar{\zeta}_0=0.22$. The proportion of chains with $\bar{\zeta}<\bar{\zeta}_0$, $P(\bar{\zeta}<\bar{\zeta}_0)$, can be obtained using \eqref{eq2p5c} on recognizing that $\bar{g}(\bar{\zeta})=Nag(\zeta)$, as follows:
\begin{equation}
P(\bar{\zeta}<\bar{\zeta}_0)=\int_0^{\bar{\zeta}_0} \frac{\bar{g}(\bar{\zeta})}{\rho_g} d\bar{\zeta}=1-\left(1- \left(\frac{\bar{\zeta}_0}{\bar{H}}\right)^{2}\right)^{3/2}=0.62
\end{equation}
For $\rho_g=0.02$, $\bar{H}=\beta_e=0.32$, hence $P(\bar{\zeta}<\bar{\zeta}_0)=0.62$. Only $62\%$ of chains satisfy the condition for $\rho_g=0.02$. For $\rho_g=0.01$, this fraction is $85\%$. So, Gaussian chain assumption is not valid for $\rho_g=0.02$ and only lower graft densities may follow the assumption. Based on \fref{monomer_density_Zparam}, we can conclude that it is valid for $\rho_g\le0.01$.

\begin{table}[!h]
	\caption{\label{beta_stretch} Stretching and extension parameters for brushes with different graft densities are listed. Large value of stretching parameter suggests strong stretching, and hence SST is applicable. However, the gaussian chain assumption is acceptable only if the extension in chains is less $1/3$, limiting validity of SST to graft densities less than $0.03$. Note that strong stretching can be achieved by increasing $N$, however extension is not affected by a change in $N$ in a strongly stretched brush.}
	\center{
	\begin{tabular}{llll}
		\hline
		 $\rho_g$ & $N$ & Stretching & Extension \\ 
		  & & ($\beta_s=\frac{3}{2}\frac{H^2}{Na^2})$ & ($\beta_e=\frac{H}{Na}$) \\ \hline
		 $0.005$ & $500$ & $31.22$ & $0.20$  \\  
		 $0.006$ & $500$ & $34.25$ & $0.21$  \\ 
		 $0.008$ & $500$ & $40.58$ & $0.23$  \\  
		 $0.01$ & $300$ & $30.34$ & $0.26$  \\  
		 $0.02$ & $300$ & $45.63$ & $0.32$  \\  
		 $0.03$ & $200$ & $40.87$ & $0.37$  \\  
		 $0.05$ & $200$ & $57.05$ & $0.44$  \\  
		 $0.06$ & $200$ & $64.48$ & $0.46$  \\ 
		 $0.08$ & $200$ & $78.57$ & $0.51$  \\ 
		  $0.1$ & $200$ & $91.95$ & $0.55$  \\  
		 $0.2$ & $200$ & $153.52$ & $0.72$  \\  
		 $0.4$ & $100$ & $133.63$ & $0.94$  \\  
		 $0.5$ & $100$ & $158.42$ & $1.03$  \\  
		 \hline
	\end{tabular}}
\end{table}

\subsection{End density} \label{EndDen_result}

\begin{figure}[!h]
	\center
	\hspace{-0.9cm}
	\includegraphics[width=8.7cm]{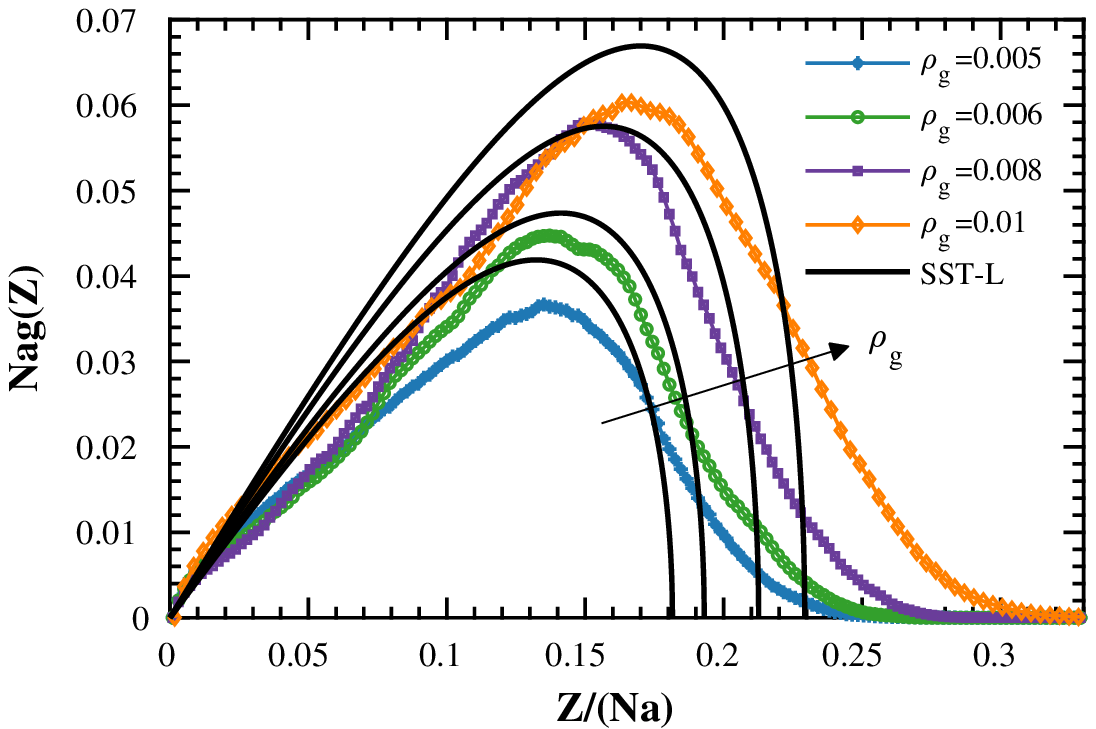}\\
	\includegraphics[width=8.7cm]{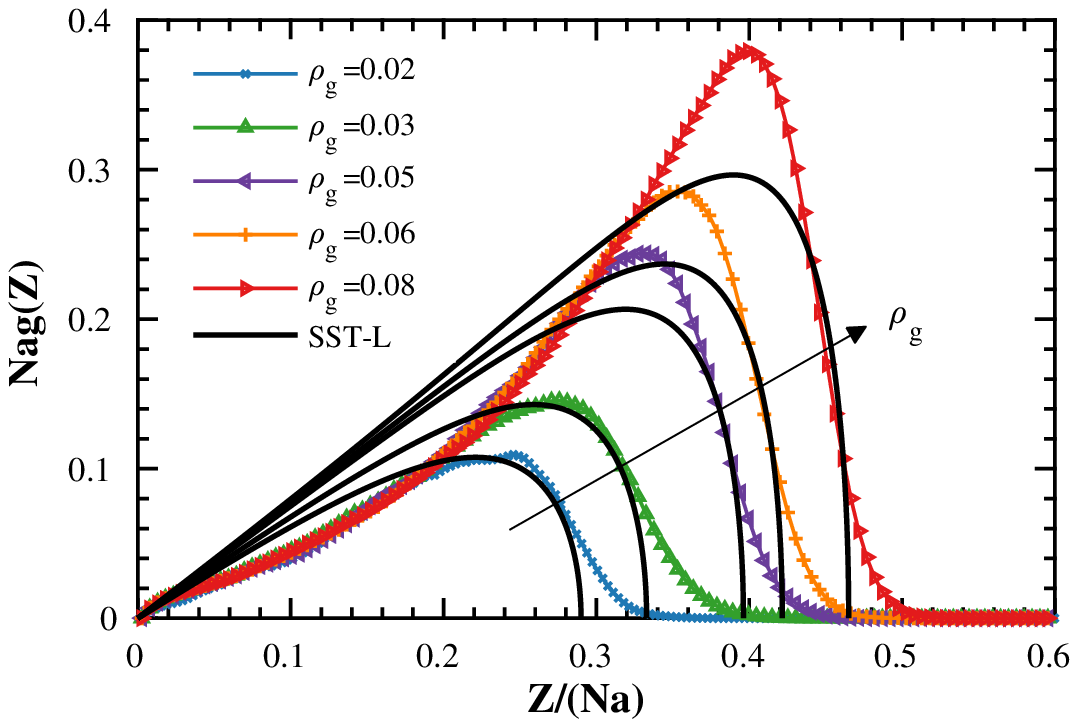}\\
	\includegraphics[width=8.7cm]{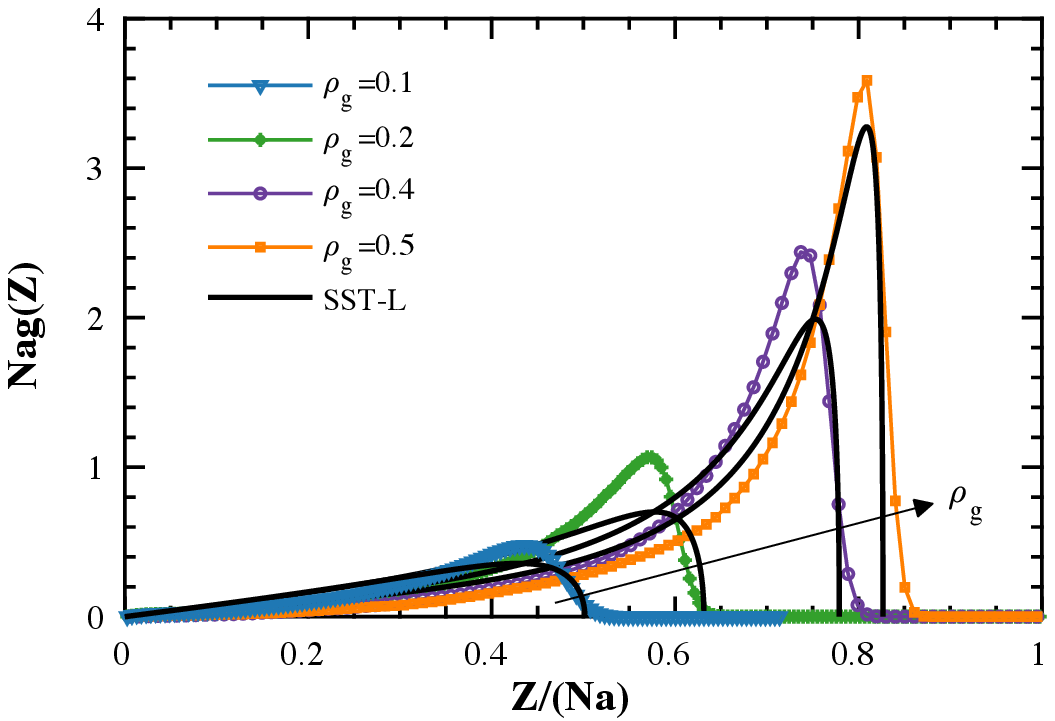}
	\caption{Variation of end density in a polymer brush with distance from the grafting surface. Prediction of end density profile from SST-L for lower graft densities (top) is very different from MD result, however we observe a better match at high graft densities (bottom).}
	\label{EndDenParam_Zparam}
\end{figure}

We plot the variation of the scaled end density of monomers with scaled distance from the grafting surface in \fref{EndDenParam_Zparam}, obtained from the MD simulations and SST-L. We observe SST-L prediction deviates considerably from the MD prediction for low graft density brushes. Generally, the curves from MD show sharper peaks and a smooth transition to zero at the brush end than those predicted by SST-L. However, with increasing graft density, we obtain a better agreement as is depicted in the last plot in Fig. \ref{EndDenParam_Zparam}. The difference at lower graft densities is related to the small value of $\beta_s$ (see Table~\ref{beta_stretch}), which results in large depletion layer and tail. Also, brush free ends increasingly concentrate to the end of the brush, as assumed in scaling theory, when graft density is very high.

\begin{figure}[!h]
	\center
	\includegraphics[width=8.7cm]{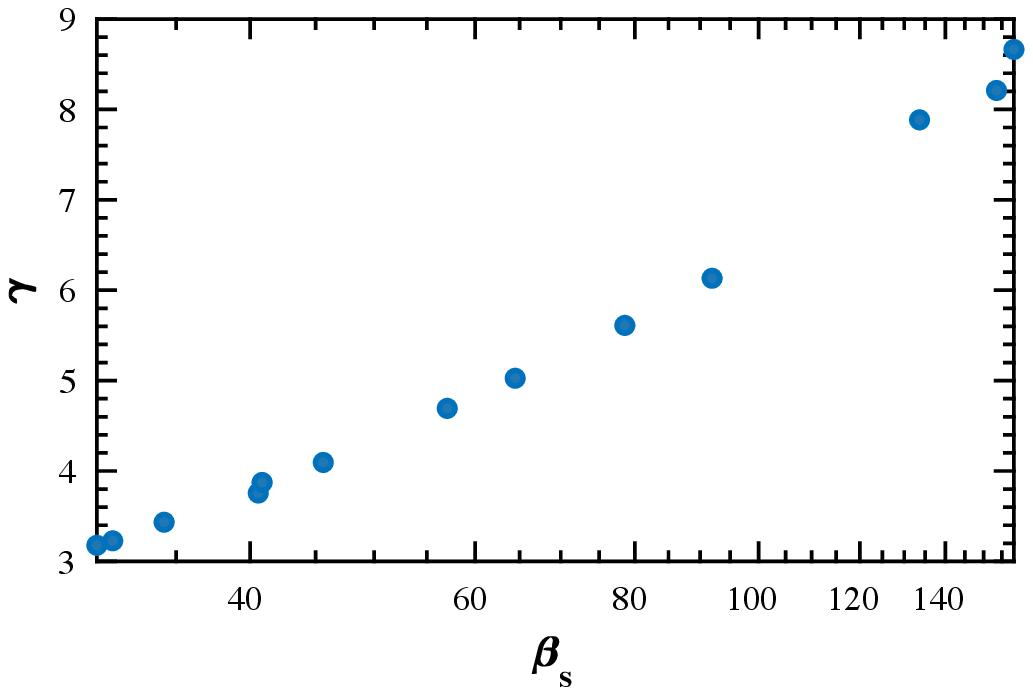}
	\caption{Effective stretching ratio ($\gamma$) is plotted vs stretching ratio ($\beta_s$) and it follows the increasing pattern observed in \cite{Siedel}. However, unlike \cite{Siedel}, we do not have points for lower $\beta_s$ in the plot.} 
	\label{gammaVSbeta}
\end{figure}

We also plot effective stretching ratio $\gamma$, defined as $\left< \zeta \right>/\left< \zeta \right>_0$, where $\left< \zeta \right>$ is mean chain end height in a brush and $\left< \zeta \right>_0$ is mean end height of a single polymer chain with no neighbouring chains as a function of $\beta_s$ in \fref{gammaVSbeta}. We observe that it follows the pattern suggested in \cite{Siedel}, however since $\beta_s$ is large in our plot, we do not see the lower end of the plot as in \cite{Siedel}. Note that in our calculations, to find $\left< \zeta \right>_0$, we assume polymer chain to be ideal, in which case $\left< \zeta \right>_0=\sqrt{2/3N}a$ \cite{Siedel}.

\subsection{Stress} \label{stress_result}
\begin{figure}[!h]
	\center
	\includegraphics[width=8.7cm]{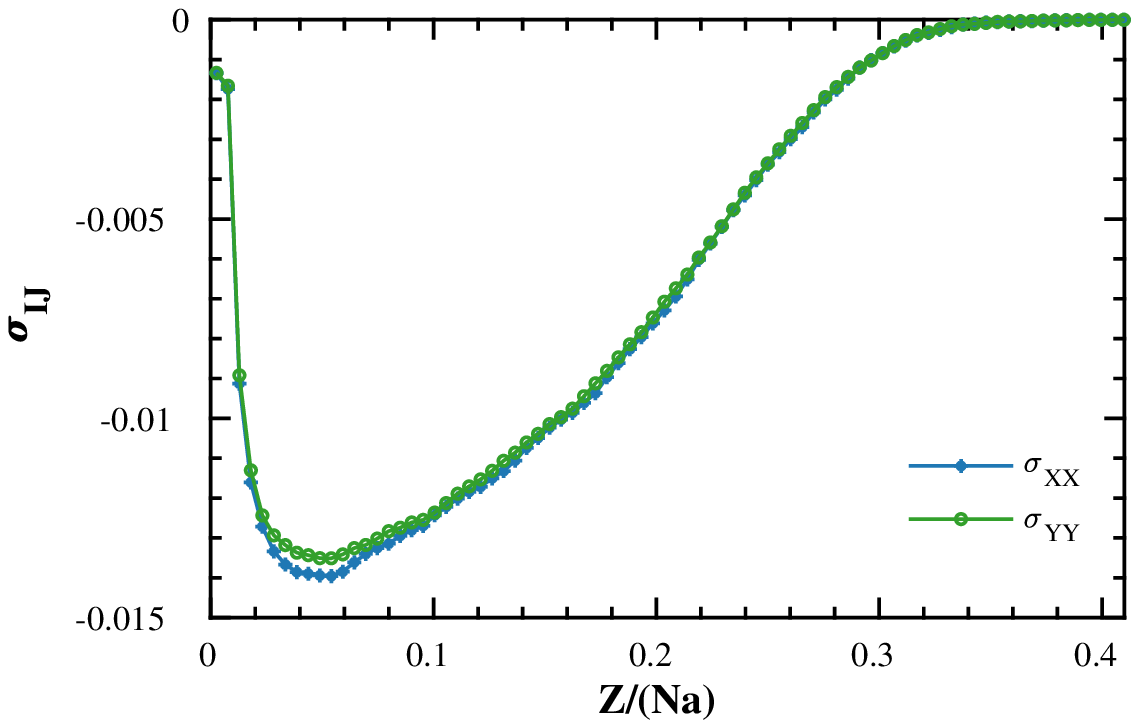}\\
	\hspace{1.1cm}
	\includegraphics[width=8.7cm]{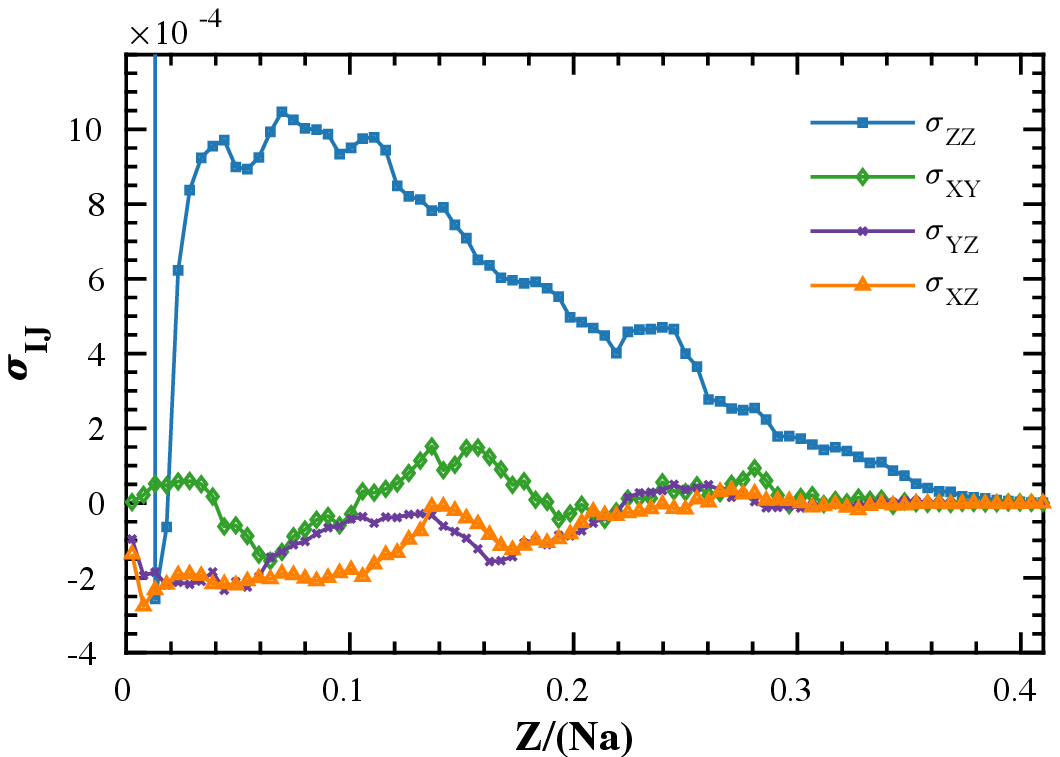}\\
	\caption{Typical stress profile in a polymer brush. The magnitudes of shear stresses are an order of magnitude smaller than the magnitude of $\sigma_{\textsc{xx}}$ and $\sigma_{\textsc{yy}}$ and hence is neglected. $\sigma_{\textsc{zz}}$ is up to $\sim 1/3$ of $\sigma_{\textsc{xx}}$ in the lowest graft density brush and decreases to less than $1\%$ for the highest graft density simulated. Also, note that near the grafting surface, $\sigma_{\textsc{zz}}$ has a very large magnitude ($0.28~\epsilon/\sigma^3$, not shown in the plot) due to wall repulsion.}
	\label{stress_Z}
\end{figure}

After validating our simulations and presenting a detailed study of the structural properties of a brush in the previous sections, we now consider the variation of stress in the polymer brush using MD and theory. The typical variation of the components of the virial stress in a brush obtained as described in Section \ref{StressMD} are shown in \fref{stress_Z} (the plots are for $\rho_g=0.03$). Notice that the normal stresses in $\textsc{x}$ and $\textsc{y}$ directions are the same as expected from symmetry among the two directions. Also, shear stresses are an order of magnitude smaller compared to normal stress in $\textsc{x}$ and $\textsc{y}$ directions and thus, they are neglected. Normal stress in $\textsc{z}$ direction is found to be up to one third of normal stress in the $\textsc{x}$ direction for the lowest graft density and the fraction decreases with increasing graft density to become less than $1\%$ for the highest graft density. This likely results from the fact that brush is not very strongly stretched at low graft densities.

\begin{figure}[!h]
	\center
	\includegraphics[width=8.7cm]{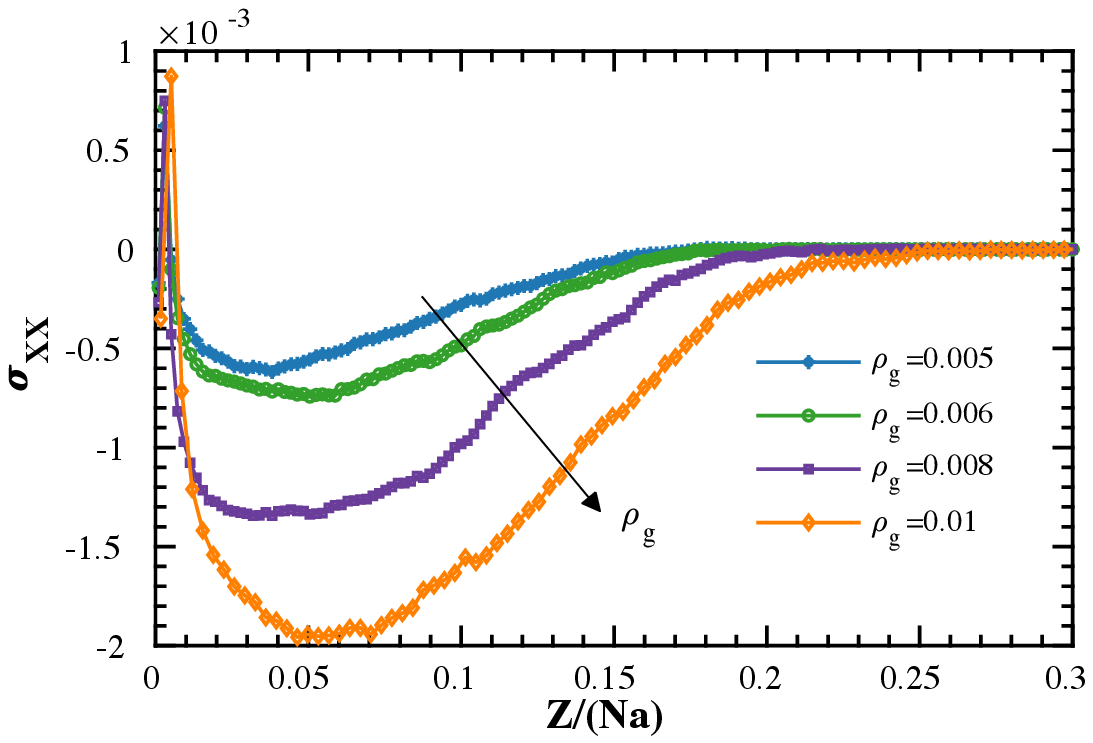}\\
	\hspace{-0.8cm}
	\includegraphics[width=8.7cm]{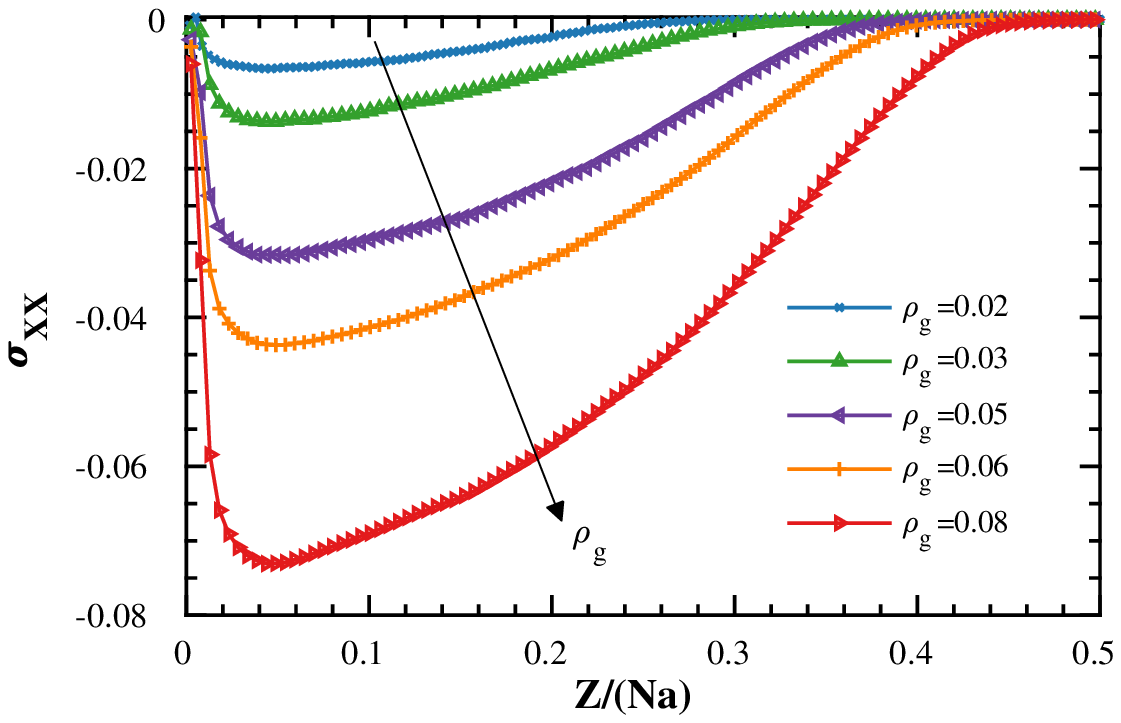}\\
	\hspace{0.1cm}
	\includegraphics[width=8.7cm]{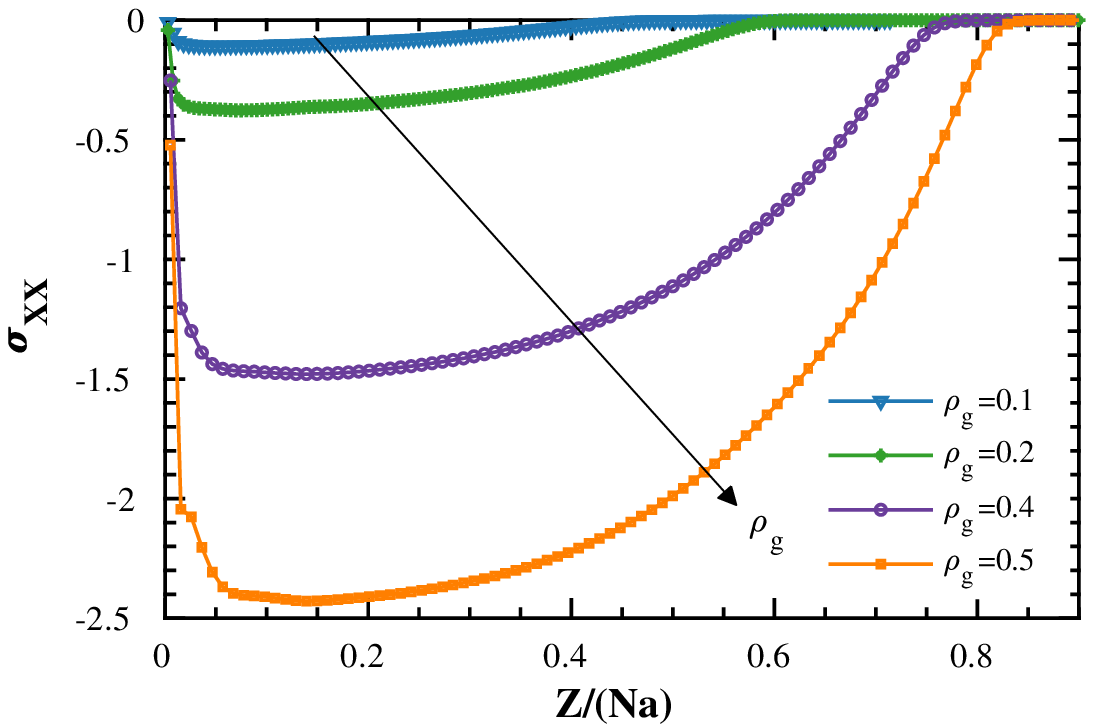}
	\caption{Variation of stress component $\sigma_{XX}$ in a polymer brush with distance from the grafting surface. Comparisons with SST-G and SST-L are shown in~\fref{Sxx_Zparam} and in~\fref{Sxxparam_Zparam}, respectively, for appropriate ranges of graft densities.}
	\label{Sxx_Z}
\end{figure}

\begin{figure}[!h]
	\center
	\includegraphics[width=8.7cm]{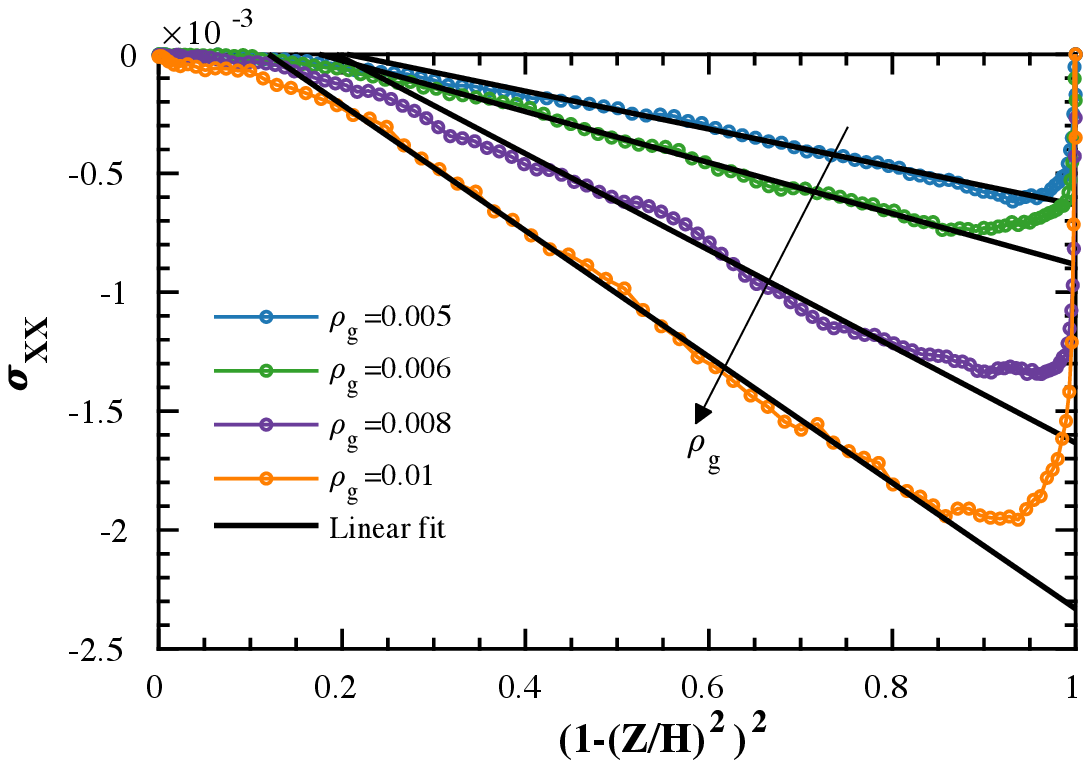}\\
	\includegraphics[width=8.6cm]{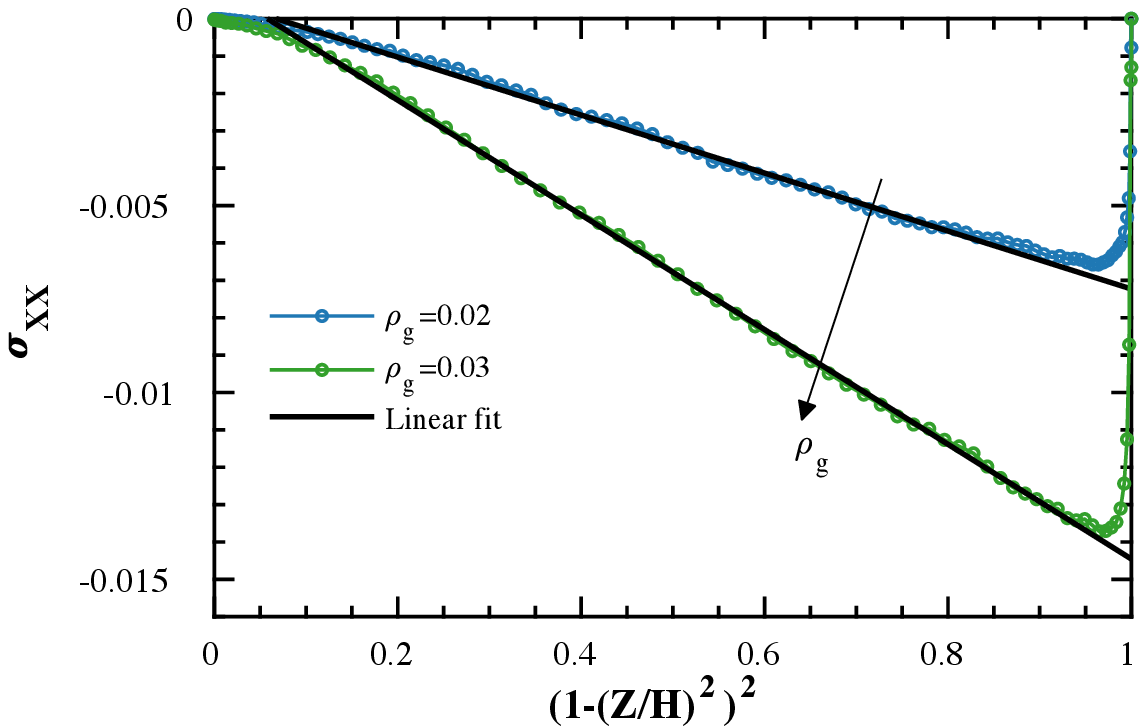}
	\caption{The plot shows variation of $\sigma_{\textsc{xx}}$ with a quartic function of the distance from the grafting surface. Simulation results show good agreement with the quartic variation prediction from SST-G for small grafting densities. Note that even though finite extensibility effects cause deviation from parabolic monomer density profile for $\rho_g=0.02, 0.03$, stress still shows quartic variation.}
	\label{Sxx_Zparam}
\end{figure}

\fref{Sxx_Z} shows the stress ($\sigma_{\textsc{xx}}$) variation in a brush as graft density is varied. Again, we distinguish low, intermediate and high graft densities. To check the validity of quartic variation of stress in low graft density brushes, as predicted in \cite{manav2018}, we plot $\sigma_{\textsc{xx}}$ as a quartic function of $\textsc{z}$ in \fref{Sxx_Zparam}. The stress profile indeed shows quartic variation within the bulk of the brush for graft densities up to $\rho_g=0.03$. At the grafted and the free ends of the brush, variation from the quartic profile is observed due to a depletion layer and a tail, respectively. Furthermore, even though we find that monomer density profile shows parabolic profile, as predicted by SST-G, for $\rho_g\le 0.01$, the quartic stress profile (also predicted by SST-G) persists up to $\rho_g=0.03$. This numerical evidence, obtained with MD simulations, validates the previous theoretical results about stress variation obtained using SST-G \cite{manav2018}.

\begin{figure}[!h]
	\center
	\includegraphics[width=8.7cm]{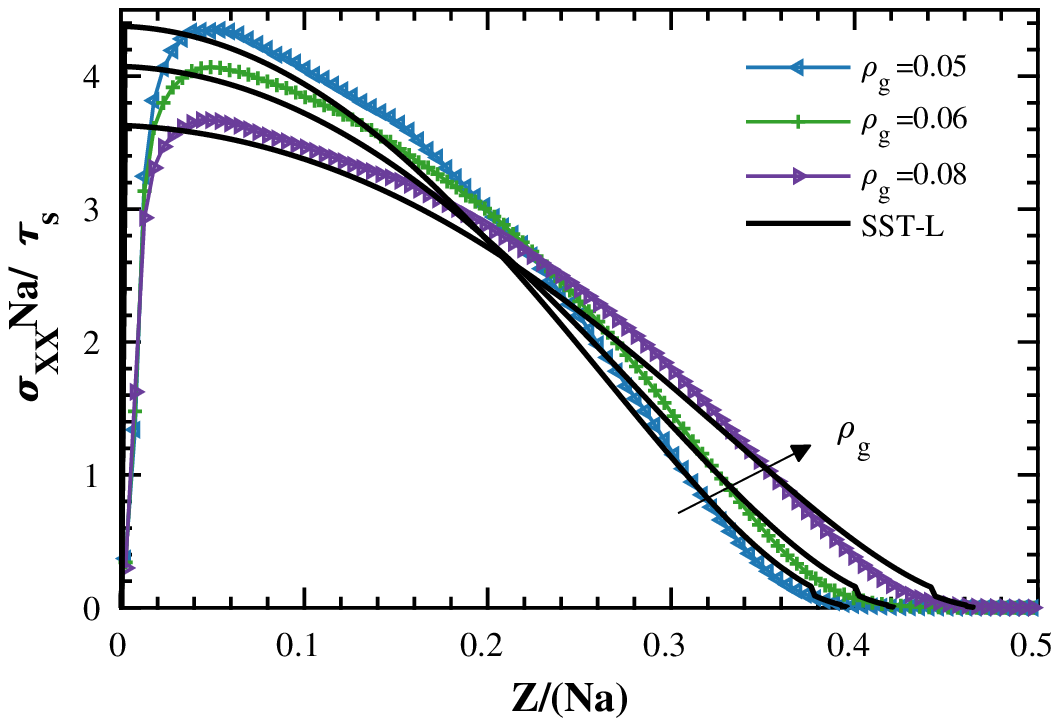}\\
	\includegraphics[width=8.7cm]{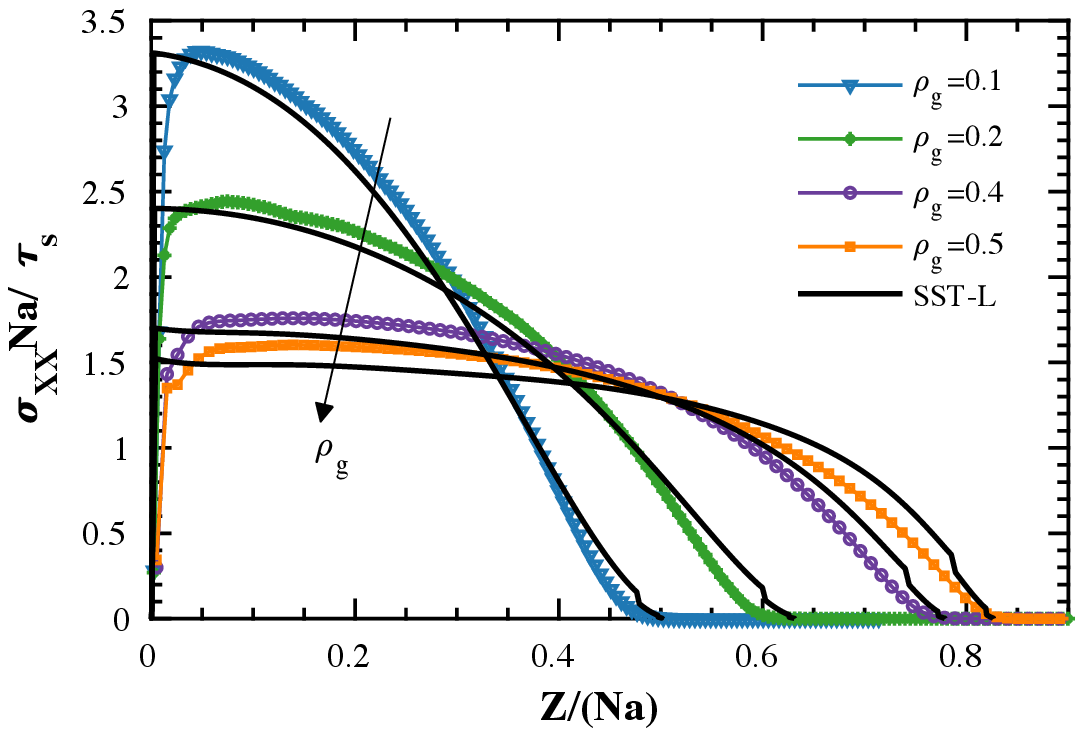}
	\caption{The plot shows the variation of normalized stress $\sigma_{\textsc{xx}}$ with the distance from the grafting surface. Simulation results show good match with the prediction from the SST-L for high graft densities.}
	\label{Sxxparam_Zparam}
\end{figure}

For higher graft densities, SST-G theory eventually breaks down and hence, we have to rely on SST-L to find stress profile. \fref{Sxxparam_Zparam} compares stress profile obtained from MD with the SST-L prediction. We find a good agreement between them. The agreement improves with increasing graft density, as monomer density and end densities are closely predicted by SST-L at high graft densities. Note that $\textsc{y}$-axis in the plots is stress divided by $\tau_s/(Na)$, where $\tau_s=\int_0^H\sigma_{\textsc{xx}}d\textsc{z}$ is surface stress, the resultant of stress in a brush. This normalization helps separate the magnitude part of the stress from the stress variation profile and we find that stress variation profile is well predicted by SST-L. Also, the stress variation curve (obtained from SST-L) near the free end of the brush has points where the curve is not smooth. As explained in Section~\ref{deriv_calc}, this is a numerical issue due to a sharp fall in the end density profile near the top of a brush.

\begin{figure}[!h]
	\center
	\includegraphics[width=8.7cm]{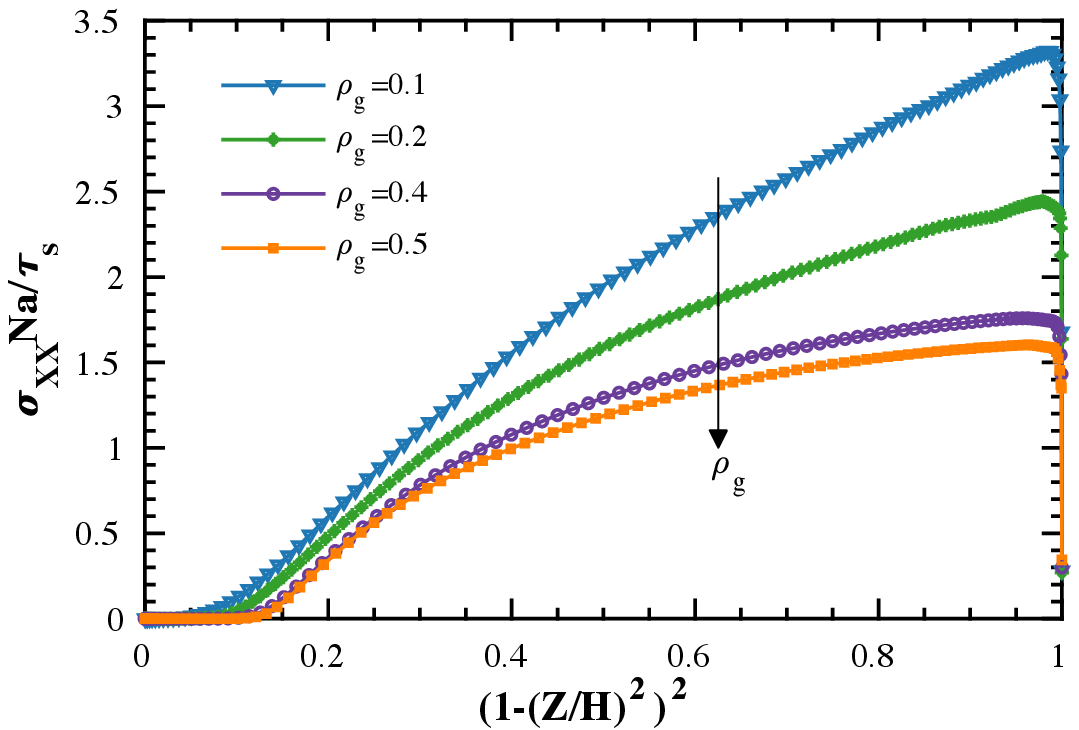}
	\caption{The plot shows variation of normalized stress $\sigma_{\textsc{xx}}$ with a quartic function of the distance from the grafting surface for very high graft densities. Notice that the stress profile appears to have a bilinear profile.}
	\label{Sxxparam_Zparam_regime3}
\end{figure}

Remarkably, for very high graft densities, the stress obtained with MD simulations suggests a \emph{bilinear} profile when plotted against $(1-(\textsc{z}/H)^2)^2$ as seen in \fref{Sxxparam_Zparam_regime3}. This suggest two regions where the polymer chain has different local stretching, which ultimately has impact on the free energy density and therefore, on stress.

\begin{figure}[!h]
	\center
	\includegraphics[width=8.7cm]{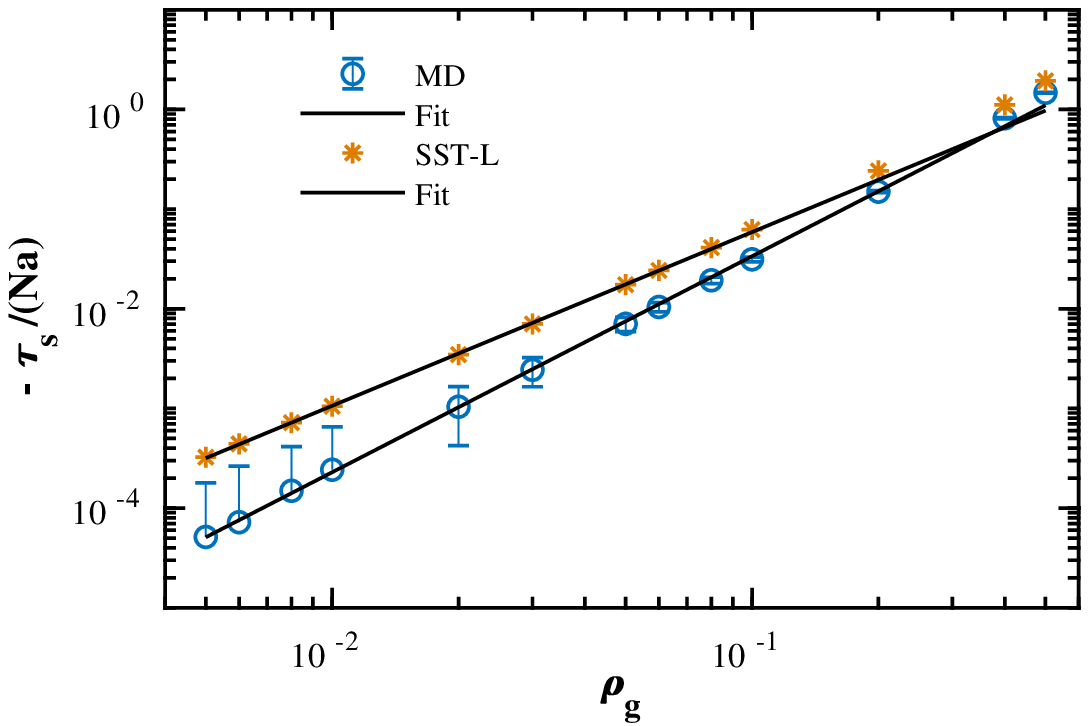}
	\caption{Variation of surface stress with respect to graft density. For the linear fit, data only up to $\rho_g=0.08$ has been used. The scaling exponent for MD data is $2.1350\pm0.0320$ as opposed to $1.7430\pm0.0160$ for SST-L. SST-G predicts this exponent to be $1.6667$, whereas scaling theory predicts it to be $1.8333$. Also, MD and SST-L both predict increasing scaling exponent for $\rho_g>0.1$.}
	\label{surfS_rho_g}
\end{figure}
Finally, we plot the dependence of the resultant surface stress ($\tau_s$) with  graft density ($\rho_g$) in \fref{surfS_rho_g}.
Considering the fact that the SST-L does not use any fitting parameter to predict the surface stress, we find that it closely predicts the magnitude of the surface stress. For high graft densities, the SST-L values match well with the MD data, while for low graft densities, more deviations are observed. This is attributed to the difference in prediction of monomer density (see Figure \ref{monomer_density_Z}) as well as end density at lower graft densities (Fig. \ref{EndDenParam_Zparam}). The scaling of surface stress with respect to graft density are, however, very different. Scaling exponent of surface stress with respect to graft density is $2.1350\pm0.0320$ from MD simulations and $1.7430\pm0.0160$ from SST-L on fitting the surface stress up to $\rho_g=0.08$. Below, we discuss the deficiencies of different methods to understand this discrepancy.

Table~\ref{scaling_comparison} compares scaling of height, free energy and stress obtained from different theories and from our simulations. Scaling of height with respect to graft density closely matches in all the theories as well in computation. In contrast, scaling of free energy and stress with respect to graft density are different in different theories. As discussed in \cite{manav2018}, mean field Flory theory inaccurately assumes that chain ends are concentrated to the free end of the brush and hence all the chains are equally and uniformly stretched. Also, it does not accounts for excluded volume correlations which occur in the limit of strong excluded volume interactions. These shortcomings lead to higher free energy predictions for a brush. Scaling theory correctly accounts for excluded volume correlations, although it also assumes that chain ends are concentrated to the free end of the brush and hence all the chains are equally stretched. Additionally, neither of the two theories account for finite extensibility of chains. SST-G does not assume equal stretching of chains and hence chain ends are distributed throughout the brush, leading to parabolic monomer density profile. But it does not account for finite extensibility of chains which is present in SST-L. However, neither SST-G nor SST-L account for excluded volume correlations, leading to over prediction of free energy as well as stress. MD simulations do not have these restrictions in principle (see Table~\ref{limitation_SST}), and give a higher scaling exponent of graft density in the expression for stress. However, it should be noted that MD simulation results are sensitive to stretching parameter $\beta_s$. For very small graft densities, $\beta_s$ is not very high, which may affect the scaling exponent.

\begin{table}[!h]
	\caption{\label{scaling_comparison} Comparison of expressions for the height, free energy and surface stress in a brush in a good solvent obtained from scaling theory, mean field Flory theory, SST-G, SST-L, and MD. Free energy and stress are in $k_B T$. Note that the scaling exponent of $\rho_g$ in the expressions for height is the same from all the theories and closely matched by semi-analytical calculation and MD on fitting the data up to $\rho_g=0.08$. The same is not true for free energy. Scaling of surface stress with respect to graft density is also different in different theories.}
	\center
	\resizebox{\linewidth}{!}{
	\begin{tabular}{llll}
		\hline
		 Method & Height & Free energy & Surface stress \\ \hline
		Scaling theory & $\sim\rho_g^{1/3}a^{5/3}N$ & $\sim \frac{5}{2}v^{1/3}\rho_g^{11/6}a^{2/3}N$ & $\sim -\frac{1}{3}v^{1/3}\rho_g^{11/6}a^{2/3}N$ \\ 
		Mean field Flory theory & $\left(\frac{1}{6}\right)^{1/3}v^{1/3}\rho_g^{1/3} a^{2/3}N$ & $\frac{9}{2}\left(\frac{1}{6}\right)^{2/3}v^{2/3}\rho_g^{10/6}a^{-2/3}N$ & $-3\left(\frac{1}{6}\right)^{2/3}v^{2/3}\rho_g^{10/6}a^{-2/3}N$ \\  
		SST-G & $\left(\frac{4}{\pi^2}\right)^{1/3}v^{1/3}\rho_g^{1/3} a^{2/3}N$ & $\frac{9}{10}\left(\frac{\pi^2}{4}\right)^{1/3}v^{2/3}\rho_g^{10/6}a^{-2/3}N$ & $-\frac{3}{5}\left(\frac{\pi^2}{4}\right)^{1/3}v^{2/3}\rho_g^{10/6}a^{-2/3}N$ \\  
		SST-L & $\sim\rho_g^{1.02/3}$ & $\sim\rho_g^{10.24/6}$ & $\sim\rho_g^{(10.46\pm 0.1)/6}$ \\  		MD & $\sim\rho_g^{1/3}$ & - & $\sim\rho_g^{(12.81\pm 0.2)/6}$ \\  
		\hline
	\end{tabular}
	}
\end{table}

%%%%%%%%%%%%%%%%%%%%%%%%%%%%%%%%%%%%%%%%%%%
\section{Conclusion}
\label{conclusion}
Stresses in a polymer brush is studied in this work using mean field theories and MD simulations. The conclusions are as follows.
\begin{enumerate}
\item Molecular dynamics simulations verify the quartic stress profile prediction of SST-G from our earlier work~\cite{manav2018}, in the low graft density regime.  Gaussian elasticity assumption is valid in this range due to small extensions, as quantified by $\beta_e$. The agreement between simulations and SST-G prediction is within the bulk of the brush and away from depletion layer and tail. Our simulations also confirm the parabolic monomer density profile.
\item Gaussian elasticity of chains breaks down at higher graft densities and has lead to discrepancies between the SST-G and MD results. This motivated the advancement of a semi-analytical parameter free theory (SST-L) based on Langevin elasticity of polymer chains, which accounts for the divergence in force-extension relation. Further, SST-L does not restrict itself to binary interactions among monomers as does SST-G. These two features are found to explain the MD results satisfactorily.
\item Prediction from SST-L for monomer density (see \fref{monomer_density_Z}) end density profile (see \fref{EndDenParam_Zparam}), brush height (see \fref{H_rho_g}), and stress profile (see \fref{Sxxparam_Zparam}) agree well with MD simulations at high graft densities. We also note that SST-L predictions for these parameters smoothly transition from SST-G at low graft densities to those of step-profile used in scaling theories at high graft densities.
\item Surface stress predicted by SST-L matches closely with MD results for high graft density. For lower graft densities, SST-L over-predicts the surface stress. Also, scaling exponents of surface stress with graft density obtained from the two theories for $\rho_g\le0.08$ are different. The precise reason for this is yet to be understood, though one can speculate about the validity of stress measures, differential end-stretching with distance from the grafting surface, MD potentials and low values of $N$.
\end{enumerate}
A natural extension of this work is to consider \emph{semiflexible} polymer chains, common in biology, based on worm-like chain models for elasticity. Also, the effect of stimuli on stress in a brush can be modelled by adding an enthalpic term to the CS equation of state, SST-L~\cite{biesheuvel2008}.  MD simulations can be extended by retaining the attraction term in pairwise interaction potential (LJ potential) to model the effect of a change in temperature, following~\cite{grest93}. These are avenues for further study and exploration.

\section{Acknowledgments}
The authors would like to thank  Natural Sciences and Engineering Research Council of Canada (NSERC) for its funding through Discovery, CREATE (NanoMat program at UBC), and the Collaborative Health Research project  jointly with the Canadian Institute of Health Research.  We thank Compute Canada for providing computational resources through its Westgrid consortium. Manav would like to thank UBC for the award of four year fellowship (4YF).

\appendix

\section{SST-L brush structure calculations}
\label{SSTLcalc}
\subsection{Calculation of $V_f$ and $\rho_g$}
To find $V_f$ at $\bar{\textsc{z}}$ in a brush of given height $\bar{H}$, first we evaluate the left hand side of \eqref{eq2p0} using the expression for $\bar{V}(\bar{\textsc{z}})$ in \eqref{eq2p11}. On the right hand side of \eqref{eq2p0}, we substitute \eqref{eq2p13}, as $\mu(\phi)=\tilde{\mu}(V_f)$ and solve for $V_f$ numerically. 

In the SST-L approach, in contrast with SST-G, height is input and not the graft density. Hence, we need to find $\rho_g$ for a given height using the following:
\begin{equation}
\rho_g=\frac{1}{A_0}\int_0^{\bar{H}}V_f d\bar{\textsc{z}}. \label{eq2p13b}
\end{equation}

\begin{figure}[!h]
\center
\includegraphics[width=8.7cm]{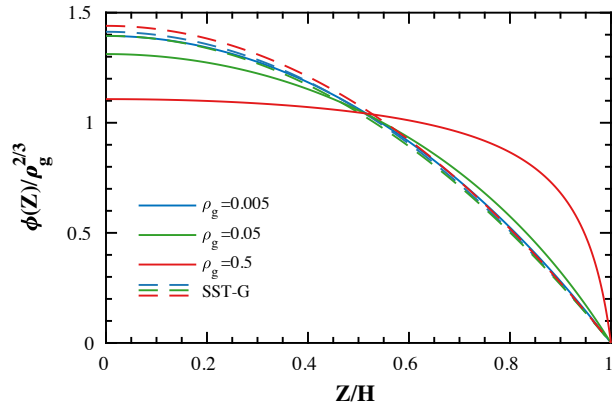}
\caption{Monomer density profile obtained from SST-L is compared with the prediction from the SST-G. Notice that SST-L (solid lines) predicts that the profile for low graft density is parabolic and matches closely with the SST-G (dashed lines) and it approaches a step profile with increasing graft density.}
\label{numden_Z_comparison}
\end{figure}

Now we can compare the monomer density profiles predicted from  SST-L and  SST-G for a brush with a given $\rho_g$, and the results are shown in~\fref{numden_Z_comparison}. To generate the plot, we make use of the relation $V_f=A_0 a \phi$ to find $\phi$. Note that, the normalized brush height ($\bar{H}$) for both the theories are prescribed to be the same for a graft density when comparing the two theories. Based on height and graft density, binary interaction parameter ($v$) in SST-G is obtained using \eqref{eq2p5b}. Using this $v$, monomer density profile is obtained using \eqref{eq2p5a}. At low graft density, the prediction for density profile from the two theories are parabolic and match closely, as expected. However, with increasing graft density, unlike SST-G, density profile predicted by SST-L, approaches step profile.

\subsection{Calculation of end density and its derivative}
\begin{figure}[!h]
\center
\includegraphics[width=8.7cm]{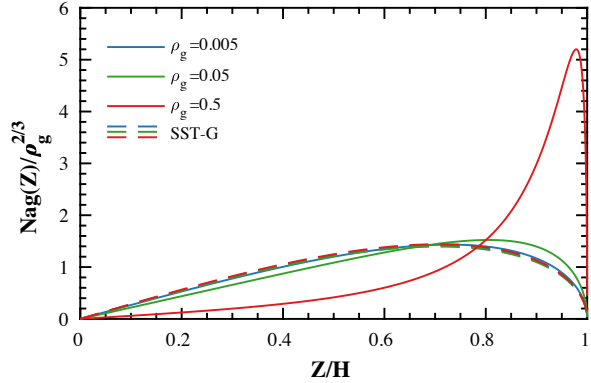}
\caption{End density profiles obtained from SST-L is compared with the prediction from SST-G. At high graft density, chain ends are predicted to lie near the free end of the brush, unlike the prediction from SST-G. }
\label{endden_Z_comparison}
\end{figure}

Normalized end density $\bar{g}(\bar{\textsc{z}})$ in a brush is obtained using the following relation (see~\cite{amoskov94,biesheuvel2008} for details):
\begin{equation}
\bar{g}(\bar{\textsc{z}})=\frac{1}{A_0}\frac{d\bar{V}(\bar{\textsc{z}})}{d\bar{\textsc{z}}}\int_0^{\bar{\textsc{z}}^*} \frac{dV_f^{\prime}}{d\nu}d\bar{\textsc{z}}^{\prime}, \label{eq2p14}
\end{equation}
where $\bar{\textsc{z}}^*$ is found from the following implicit relation:
\begin{equation}
\bar{V}(\bar{\textsc{z}}^*)=\bar{V}(\bar{H})-\bar{V}(\bar{\textsc{z}}), \label{eq2p15}
\end{equation}
and $\nu=\bar{V}(\bar{H})-\bar{V}(\bar{\textsc{z}})-\bar{V}(\bar{\textsc{z}}^{\prime})$. Finding $\frac{dV_f^{\prime}}{d\nu}$ directly is difficult. So, we first find $\frac{d\nu}{dV_f^{\prime}}$ using \eqref{eq2p1} and \eqref{eq2p13} as suggested in \cite{biesheuvel2008}:
\begin{equation}
\frac{d\nu}{dV_f^{\prime}}=\frac{1}{d}\frac{6+3V_f^{\prime}-4V_f^{{\prime}^2}+V_f^{{\prime}^3}}{(1-V_f^{\prime})^4}, \label{eq2p16}
\end{equation}
and obtain $\frac{dV_f^{\prime}}{d\nu}$ using $\frac{dV_f^{\prime}}{d\nu}=1/(\frac{d\nu}{dV_f^{\prime}})$. Note that $V_f^{\prime}\ne V_f$, and is obtained by solving the following equation for given $\bar{\textsc{z}}$ and $\bar{\textsc{z}}^{\prime}$:
\begin{equation}
\tilde{\mu}(V_f^{\prime})=\bar{V}(\bar{H})-\bar{V}(\bar{\textsc{z}})-\bar{V}(\bar{\textsc{z}}^{\prime}). \label{eq2p17}
\end{equation}
\fref{endden_Z_comparison} compares end density profiles for a few graft densities as predicted by SST-G and SST-L. Again, for low graft densities, the two predictions match very well. For high graft densities, however, most of the chain ends approach the free end of the brush.

The calculation of stress in a brush requires evaluation of $\left[\frac{\partial \bar{g}(\bar{\zeta})}{\partial \epsilon_{\textsc{xx}}}\right]_{\frac{\partial \bar{\zeta}}{\partial \epsilon_{\textsc{xx}}}=0}$ ($\bar{\zeta}$ is integration parameter). It is obtained by taking derivative of \eqref{eq2p14}:
\begin{equation}
\frac{\partial \bar{g}(\bar{\zeta})}{\partial \epsilon_{\textsc{xx}}}=\frac{1}{A_0}\frac{d\bar{V}(\bar{\zeta})}{d\bar{\zeta}}\left(\frac{\partial \bar{\zeta}^*}{\partial \epsilon_{\textsc{xx}}}\left[ \frac{dV_f^{\prime}}{d\nu} \right]_{\bar{\zeta}=\bar{\zeta}^*} +\int_0^{\bar{\zeta}^*} \frac{\partial}{\partial \epsilon_{\textsc{xx}}}\left( \frac{dV_f^{\prime}}{d\nu} \right)d\bar{\textsc{z}}^{\prime}\right),
\label{eq2p26}
\end{equation}
$\frac{d\bar{V}(\bar{\zeta})}{d\bar{\zeta}}$ is obtained using \eqref{eq2p11}. To find $\frac{\partial \bar{\zeta}^*}{\partial \epsilon_{\textsc{xx}}}$, we take derivative of \eqref{eq2p15}.
\begin{equation}
\frac{\partial \bar{\zeta}^*}{\partial \epsilon_{\textsc{xx}}}=\frac{\bar{H}}{\bar{\zeta}^*}\frac{2\bar{H}^4-4\bar{H}^2+5}{2\bar{\zeta}^{*^4}-4\bar{\zeta}^{*^2}+5}\left(\frac{1-\bar{\zeta}^{*^2}}{1-\bar{H}^2}\right)^2\frac{\partial \bar{H}}{\partial \epsilon_{\textsc{xx}}}.
\label{eq2p27}
\end{equation}
$\frac{dV_f^{\prime}}{d\nu}$ is obtained from \eqref{eq2p16} and $\frac{\partial}{\partial \epsilon_{\textsc{xx}}}\left( \frac{dV_f^{\prime}}{d\nu} \right)$ is obtained by taking derivative of \eqref{eq2p16}:
\begin{align}
\frac{\partial}{\partial \epsilon_{\textsc{xx}}}\left( \frac{dV_f^{\prime}}{d\nu} \right)&=\frac{\partial}{\partial \epsilon_{\textsc{xx}}}\left(\frac{1}{ \frac{d\nu}{dV_f^{\prime}}} \right) \nonumber \\
&=-d\left(\frac{(1-V_f^{\prime})^3(V_f^{{\prime}^3}-5V_f^{{\prime}^2}+V_f^{\prime}+27)}{(6+3V_f^{\prime}-4V_f^{{\prime}^2}+V_f^{{\prime}^3})^2}\right)\frac{\partial V_f^{\prime}}{\partial\epsilon_{\textsc{xx}}}, \label{eq2p28}
\end{align}
where
\begin{equation}
\frac{\partial V_f^{\prime}}{\partial\epsilon_{\textsc{xx}}}=\frac{8}{5}\frac{d}{a}\left(\frac{\bar{H}(2\bar{H}^4-4\bar{H}^2+5)}{(1-\bar{H}^2)^2}\right)\left(\frac{(1-V_f^{\prime})^4}{6+3V_f^{\prime}-4V_f^{{\prime}^2}+V_f^{{\prime}^3}}\right)\frac{\partial \bar{H}}{\partial \epsilon_{\textsc{xx}}}. \label{eq2p29}
\end{equation}
Since we already know $\bar{H}$, $\frac{\partial \bar{H}}{\partial \epsilon_{\textsc{xx}}}$, $V_f$ and we can numerically calculate $\bar{\zeta}^*$, all the expressions above can be numerically calculated to finally obtain $\left[\frac{\partial \bar{g}(\bar{\zeta})}{\partial \epsilon_{\textsc{xx}}}\right]_{\frac{\partial \bar{\zeta}}{\partial \epsilon_{\textsc{xx}}}=0}$.

\section{Generating initial configuration of a brush with approximately parabolic density profile}
\label{init_config}
For low graft density brushes, achieving hight $\beta_s$ to be able to make comparisons with SST requires large number of beads per chain ($N$). However, equilibrating a brush with large $N$ is very difficult due to the fact that relaxation time for a chain increases very fast with increasing $N$. So, starting from a good initial configuration is imperative. As brushes at low graft density show parabolic profile, SST-G results, summarized in Section~\ref{SST_Gaussian}, can be used to generate initial brush configuration. Below we describe the steps involved.
\begin{enumerate}
\item Determine total number of chains, $N_g$, in the brush. Define a surface and decide locations of $N_g$ grafting points. I chose equispaced grid points as grafting points.
\item Calculate a tentative brush height, $H_t$, using \eqref{eq2p5b}. Divide the region between $\textsc{z}=0-H_t$ into $n_{bins}$. Choose an optimal value of $n_{bins}$ so that $g(\zeta)$ vs $\zeta$ is close to the curve predicted by \eqref{eq2p5c}.
\item Using end probability $P_E(\zeta)=g(\zeta)/\rho_g$, where $g(\zeta)$ is found using \eqref{eq2p5c}, calculate number of chain ends in the $i^{th}$ bin as $N_{CE}(i)=round(N_g\times \int_{\textsc{z}_{low}(i)}^{\textsc{z}_{up}(i)}P_E(\zeta)d\zeta)$. Here $\textsc{z}_{low}(i)$ and $\textsc{z}_{up}(i)$ are the lower and upper boundaries of $i^{\rm th}$ bin. $N_g-\sum_{i} N_{CE}$ chains are added to the bin with the maximum $N_{CE}$.
\item Starting from the first bin, randomly assign $N_{CE}(i)$ grafting points to each bin. By doing this, we ensure that a chain starting from a given grafting point ends in a particular bin.
\item Now we start defining chains originating from each grafting location one by one. For each grafting location we already know the chain end $\zeta$. Also, we know $\frac{d\textsc{z}}{dn}=E(\textsc{z},\zeta)$ from \eqref{eq2p5d}. Using $\Delta n=1$, we find $\Delta \textsc{z}=E(\textsc{z},\zeta)$. Hence, $\textsc{z}$-coordinate of $(i+1)^{\rm th}$ bead in a given chain is given by, $\textsc{z}_{i+1}=\textsc{z}_i+E(\textsc{z}_i,\zeta)$. $\Delta \textsc{x}$ and $\Delta \textsc{y}$, such that $\textsc{x}_{i+1}=\textsc{x}_i+\Delta \textsc{x}$ and $\textsc{y}_{i+1}=\textsc{y}_i+\Delta \textsc{y}$, are randomly chosen (with the constraint that beads do not go outside the simulation box) to ensure that $\sqrt{\Delta \textsc{x}^2+\Delta \textsc{y}^2+\Delta \textsc{y}^2}$=bond length. If $\textsc{z}_{k}\ge\zeta$, we constrain  $\textsc{z}_{i}=\zeta$ for all $i\ge k$.
\end{enumerate}

\bibliography{references}

\end{document}